\documentclass{aa}
\usepackage{mathpazo}
\usepackage{graphicx}
\usepackage[T1]{fontenc}
\usepackage{amssymb}
\usepackage{pifont}
\usepackage{xcolor}
\usepackage{appendix}
\usepackage{multicol}
\usepackage[flushleft]{threeparttable}
\usepackage{array,booktabs}
\usepackage{amsmath}
\usepackage{textcomp}
\usepackage{multirow}
\usepackage{mathtools}
\usepackage{txfonts}
\usepackage{hyperref}
\hypersetup{
    colorlinks=true,
    linkcolor=blue,
    filecolor=red,
    citecolor=blue ,
    urlcolor=blue,
}

\begin{document} 

  \title{Linking ice and gas in the Serpens low-mass star-forming region}
   \author{G. Perotti
          \inst{1}\fnmsep,
          W. R. M. Rocha\inst{1},
          J. K. J{\o}rgensen\inst{1},
          L. E. Kristensen\inst{1},
          H. J. Fraser\inst{2},
          \and
          K. M. Pontoppidan\inst{3}
          }

   \institute{Niels Bohr Institute \& Centre for Star and Planet Formation, University of Copenhagen, {\O}ster Voldgade 5$-$7, 1350 Copenhagen K., Denmark\\
              \email{giulia.perotti@nbi.ku.dk}
         \and
             School of Physical Sciences, The Open University, Walton Hall, Milton Keynes, MK7 6AA, United Kingdom
         \and
             Space Telescope Science Institute, 3700 San Martin Drive, Baltimore, MD 21218, USA
             }

   \date{Received --; accepted --}

  \abstract
   {The interaction between dust, ice, and gas during the formation of stars produces complex organic molecules. While observations indicate that several species are formed on ice-covered dust grains and are released into the gas phase, the exact chemical interplay between solid and gas phases and their relative importance remain unclear.}
   {Our goal is to study the interplay between dust, ice, and gas in regions of low-mass star formation through ice- and gas-mapping and by directly measuring gas-to-ice ratios. This provides constraints on the routes that lead to the chemical complexity that is observed in solid and gas phases.}
  {We present observations of gas-phase methanol (CH$_3$OH) and carbon monoxide ($^{13}$CO and C$^{18}$O) at 1.3~mm towards ten low-mass young protostars in the Serpens SVS~4 cluster from the SubMillimeter Array (SMA) and the Atacama Pathfinder EXperiment (APEX) telescope. We used archival data from the Very Large Telescope (VLT) to derive abundances of ice H$_2$O, CO, and CH$_3$OH towards the same region. Finally, we constructed gas-ice maps of SVS~4 and directly measured CO and CH$_3$OH gas-to-ice ratios.}
  {The SVS~4 cluster is characterised by a global temperature of 15 $\pm$ 5~K. At this temperature, the chemical behaviours of CH$_3$OH and CO are anti-correlated: larger variations are observed for CH$_3$OH gas than for CH$_3$OH ice, whereas the opposite is seen for CO. The gas-to-ice ratios ($N_\mathrm{gas}$/$N_\mathrm{ice}$) range from  1~$-$~6 for CO and $\mathrm{1.4 \times 10^{-4} - 3.7 \times 10^{-3}}$ for CH$_3$OH. The CO gas-maps trace an extended gaseous component that is not sensitive to the effect of freeze-out. Because of temperature variations and dust heating around 20~K, the frozen CO is efficiently desorbed. The CH$_3$OH gas-maps, in contrast, probe regions where methanol is predominantly formed and present in ices and is released into the gas phase through non-thermal desorption mechanisms.}
    {Combining gas- and ice-mapping techniques, we measure gas-to-ice ratios of CO and CH$_3$OH in the SVS~4 cluster. The CH$_3$OH gas-to-ice ratio agrees with values that were previously reported for embedded Class 0/I low-mass protostars. We find that there is no straightforward correlation between CO and CH$_3$OH gas with their ice counterparts in the cluster. This is likely related to the complex morphology of SVS~4: the Class 0 protostar SMM~4 and its envelope lie in the vicinity, and the outflow associated with SMM~4 intersects the cluster. This study serves as a pathfinder for future observations with ALMA and the James Webb Space Telescope (JWST) that will provide high-sensitivity gas-ice maps of molecules more complex than methanol. Such comparative maps will be essential to constrain the chemical routes that regulate the chemical complexity in star-forming regions.}
  
    \titlerunning{Linking ice and gas in the Serpens low-mass star-forming region}
    \authorrunning{G. Perotti et al.}
   \maketitle


\section{Introduction}
\label{Intro}
The interplay between ice and gas in the Universe is an elusive relationship spanning from the formation of simple molecules to the creation of more complex organics; the building blocks of life in planetary systems like our own. 
In star-forming regions inside dense molecular clouds, the most likely mechanism to form complex organic molecules (COMs) is the thermal and energetic processing of ice-covered dust grains \citep{Boogert2015,Oberg2016}. Unfortunately, detecting complex icy molecules is challenging because the vibrational modes of functional groups associated with these frozen molecules are blended in the infrared. So far, methanol (CH$_3$OH) is the most complex molecule that has been securely detected in the ices in the interstellar medium, although ice features of a few other species, including acetaldehyde (CH$_3$CHO) and ethanol (CH$_3$CH$_2$OH), have also been suggested to contribute to the infrared spectra of young protostars \citep{Schutte1999, Oberg2011_spitzer, Scheltinga2018}. Gas-phase (sub)millimeter single-dish and interferometric observations have demonstrated that a plethora of molecules more complex than methanol exist in the gas phase towards low- and high-mass star-forming regions (see e.g. reviews by \citealt{Herbst2009} and \citealt{Jorgensen2020}). One way to access the complex organic inventory of molecular ices consists of indirectly deriving ice abundances from gas-phase mapping \citep{Whittet2011} and subsequently reverse-engineering their path into the gas phase to constrain their origins in the ices. However, this method relies on assumptions about how efficiently icy molecules are released into the gas. To investigate the ice-gas interaction in star-forming regions, combined gas and ice maps are required to constrain the physical and chemical environment. In this paper, gas and ice maps of methanol (CH$_3$OH) and carbon monoxide (CO) are presented. The data are used to determine CH$_3$OH and CO gas-to-ice ratios and thus to shed light onto the chemical processes occurring in cold low-mass star-forming regions.  

Methanol is one of the key species in studies of the organic content of star-forming regions: it is the precursor of many complex organic molecules \citep{Nuevo2018,Rivilla2019} and the ideal candidate to link gas and ice processes: at the low temperatures of cold environments, there is no efficient methanol formation path in the gas phase \citep{Turner1998,Garrod2007}, but laboratory experiments have demonstrated that the production of methanol can successfully occur on ices at low temperatures ($\sim$10~K) through grain-surface hydrogenation reactions of CO ice \citep{Watanabe2002,Fuchs2009,Linnartz2015}. After it has formed in the ices, the question remains at what level it is released back into the gas phase through thermal and non-thermal processes. \citet{Bertin2016} and \citet{Cruz-Diaz2016} have experimentally shown that the methanol desorption yield is at least one order of magnitude lower than the current value that is largely used in the literature and was measured by \citet{Oberg2009a}, which causes the methanol gas-phase abundance to be overestimated in astrochemical models. Theoretical three-phase chemical models that combine gas-phase chemistry with bulk and surface have addressed the effect of the reactive desorption \citep{Dulieu2013} on the gas-phase abundance of molecules \citep{Cazaux2016}. In these models, neither photons nor cosmic rays play a role in the desorption mechanism, and the molecules are released into the gas phase upon formation on grains. \citet{Cazaux2016} showed that the gas-to-ice ratio for methanol is $\sim$~10$^{-4}$ at 10$^5$ years when the ice desorption is taken into account. In the opposite case, when the icy molecules are not released into the gas phase, the methanol gas-to-ice ratio is $\sim$~10$^{-6}$. 

This prediction can be tested observationally by combining gas-phase and ice observations to directly measure the gas-to-ice ratios and desorption efficiency. Maps like this were created for carbon monoxide by \citet{Noble2017} for B~35A, a star-forming dense core in Orion. The authors used data from the AKARI satellite and ground-based observations from the Institut de RadioAstronomie Millim\'etrique (IRAM) 30 m telescope and the James Clerk Maxwell Telescope (JCMT). \citet{Noble2017} concluded that there is no trivial relationship between ice and gas. No obvious trends were found between CO ice and gas-phase C$^{18}$O, suggesting that the interplay between CO freeze-out and desorption is complex in this region. Because methanol is the best candidate to bridge ice and gas chemistries, we have targeted the SVS~4 cluster, which currently holds the largest reservoir of ice methanol in the nearby star-forming regions \citep{Pontoppidan2004}.

\begin{figure}
\centering
\includegraphics[width=3.75in]{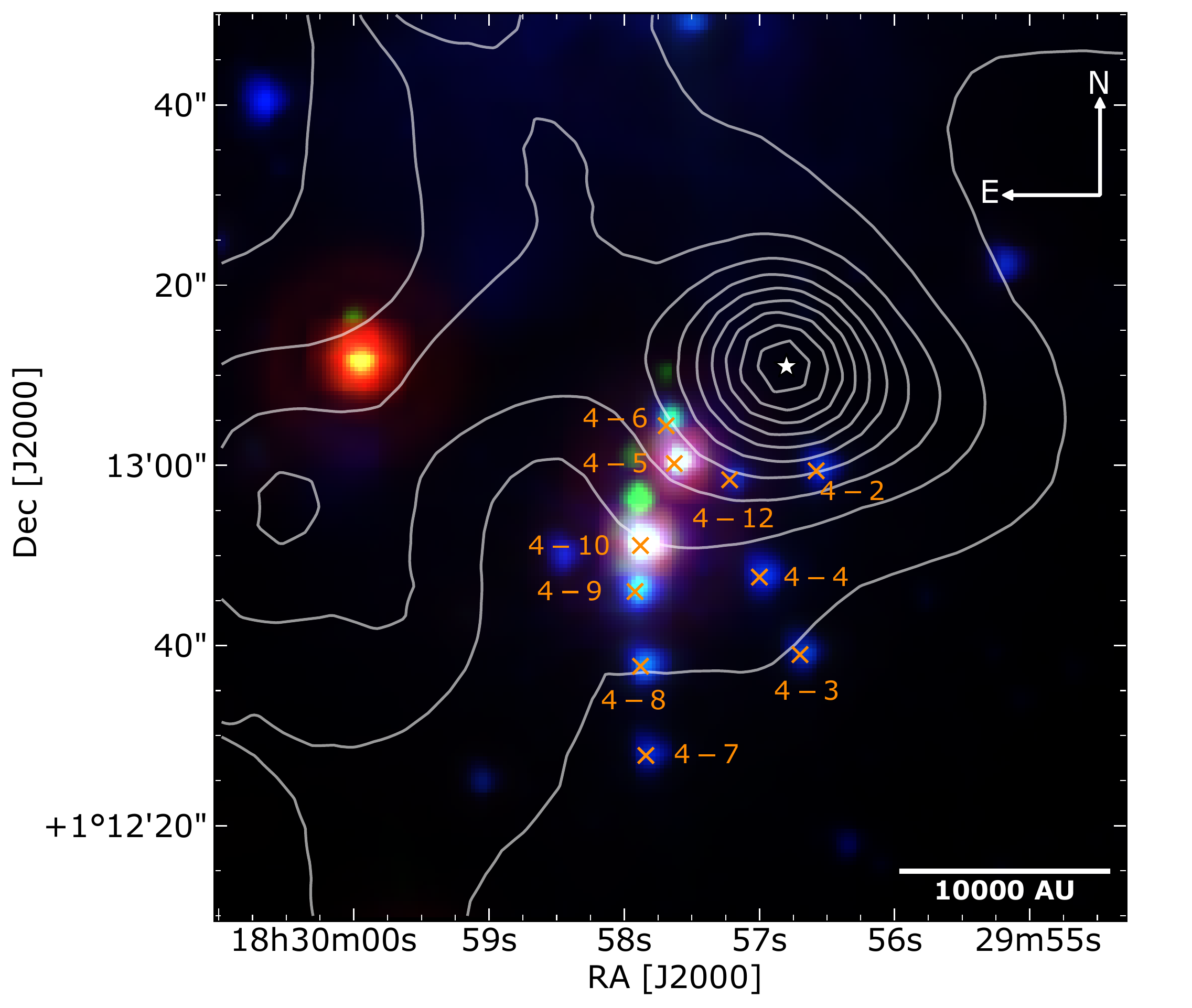}
\caption{Three-colour image of the SVS~4 cluster overlaid with SCUBA-2 850~$\mu$m density flux in mJy beam$^{-1}$ \citep{Herczeg2017}. The contours are in decreasing steps of 10\% starting at 3120 mJy beam$^{-1}$. The composite is made from \textit{Spitzer}~IRAC 3.6~$\mu$m (blue), 8.0~$\mu$m (green), and MIPS 24.0~$\mu$m (red) bands \citep{Fazio2004}. The dark orange crosses mark the position of the SVS~4 sources, and the white star represents SMM~4.}
\label{rgb_Serpens}
\end{figure}

SVS~4 is a small dense cluster located to the south-east of the Serpens molecular cloud. With 11 low- to intermediate-mass young stellar objects (YSOs) within a region of 20 000~AU, it is one of the densest YSO clusters ever observed \citep{Eiroa1989, Pontoppidan2003b, Eiroa2008}. The SVS~4 young stars are all situated within $\sim$~20$\arcsec$ at most from one another, as shown in Figure~\ref{rgb_Serpens} and listed in Table~\ref{table:samples_sources}. Additionally, the Class 0 protostar SMM~4, which has an outflow and a large envelope, is located only 26$\arcsec$ ($\sim$10 000~AU) from the centre of the cluster. According to \citet{Pontoppidan2004}, the cluster is located inside the south-eastern part of the envelope of SMM~4. As a result, lines of sight towards the SVS~4 stars probe the envelope material and the outflow of the Class 0 protostar. Among the SVS~4 cluster members, SVS~4$-$9 is an intermediate-mass YSO, and \citet{Preibisch2003} found it to be the brightest X-ray source in the Serpens molecular cloud. SVS~4$-$3 and SVS~4$-$6 are also X-ray sources \citep{Preibisch2004, Giardino2007}.

The ice content of the SVS~4 region has been investigated extensively by \citet{Pontoppidan2003a, Pontoppidan2003b, Pontoppidan2004, Pontoppidan2008}. These studies found that the SVS~4 cluster had one of the highest methanol ice abundances of low-mass stars (up to 28\% relative to water ice and 3~$\times$~10$^{-5}$ compared to gas-phase H$_2$) reported to date \citep{Pontoppidan2003a, Pontoppidan2004}. The SVS~4 methanol ice abundance relative to water is comparable to the one observed towards the most methanol-rich massive stars known \citep{Pontoppidan2003a}.
The nature of the Class~0 protostar SMM~4 itself has been extensively studied at millimeter and sub-millimeter wavelengths. These studies have revealed that the continuum emission towards SMM~4 is spatially resolved into two Class 0 protostars, SMM~4A and SMM~4B, embedded in the same core \citep{Lee2014, Aso2018, Maury2019}. In contrast, little has been done to investigate the gas-phase molecular content of the outer envelope of SMM~4, where the SVS~4 stars are located. \citet{Oberg2009a} observed methanol towards SVS~4$-$5 in the $2_K-1_K$ rotational band with the IRAM 30~m telescope providing a column density of $\mathrm{2.7 \times 10^{14} \; cm^{-2}}$ for a rotational temperature of about 8~K. \citet{Oberg2009a} inferred that the spectrum of SVS~4$-$5 shows multiple emission components, which reflects the complexity of the cluster and also the contributions from the outflow associated with SMM~4.
\citet{Kristensen2010} mapped the gas-phase methanol in the $7_K-6_K$ rotational band with JCMT towards SMM~4 and two outflow positions (SMM~4$-$W and SMM~4$-$S), and found a variation between $\mathrm{10^{14} - 10^{15} \; cm^{-2}}$ that was attributed to the methanol sputtering triggered by the outflow. 
Other than methanol, \citet{Oberg2011a} observed a suite of molecular transitions for the complex organic O-bearing species HCOOCH$_3$, CH$_3$CHO, and CH$_3$OCH$_3$ towards SMM~4 and SMM~4$-$W in 17$\arcsec$ and 28$\arcsec$ beams with IRAM, indicating a rich chemistry in the vicinity of SMM~4 and the low-mass outflow SMM~4$-$W. 

In this paper we present millimeter interferometric and single-dish gas observations of methanol ($J$= $5_K-4_K$) and carbon monoxide ($^{13}$CO and C$^{18}$O, $J$= $2-1$) towards the Serpens SVS~4 cluster. In a complementary manner, we analyse L- and M-band infrared ice spectra of ten SVS~4 stars taken from the Very Large Telescope (VLT-ISAAC) archive\footnote{http://archive.eso.org/wdb/wdb/eso/isaac/form}. 
We combine gas- and ice-mapping techniques and directly measure CO and CH$_3$OH gas-to-ice ratios in the cluster. The resulting gas-ice maps and ratios serve to constrain the routes leading to chemical complexity in solid and gas phases of cold low-mass star-forming regions.

The paper is structured as follows. In Section~\ref{observations} we present the ice and gas-phase observations together with the archival data. Section \ref{results} summarises the method we adopted to derive the ice and gas column densities and reports the observational results. Section~\ref{analysis} interprets the gas-ice maps and analyses the gas and ice variations. Section~\ref{discussion} discusses the results and provides a comparison between observed and predicted CO and CH$_3$OH gas-to-ice ratios. Finally, Section~\ref{conclusions} concludes the paper. 


\section{Observations and archival data}
\label{observations}

\subsection{SMA and APEX observations}
\label{SMA_APEX observations}
The SVS~4 cluster was observed on May 13, 2017, using the eight-antenna SubMillimeter Array (SMA; \citealt{Ho2004}). The array was in its subcompact configuration, resulting in baselines between 8$-$45~m. The region was covered by two pointings whose dimensions are dictated by the SMA primary beam size of 50$''$. The first pointing was centred on the Class 0 protostar Serpens SMM~4, and the second was offset by one half SMA primary beam to the south-east. The exact coordinates of the two SMA pointings are $\alpha_{J2000} = 18^\mathrm{h}29^\mathrm{m}57^\mathrm{s}.63$, $\delta_{J2000} = +01^\circ 13\arcmin 00\farcs2,$ and $\alpha_{J2000} = 18^\mathrm{h}29^\mathrm{m}57^\mathrm{s}.88$, $\delta_{J2000}=+01^\circ 12\arcmin 37\farcs7$. 
The pointings covered frequencies ranging from 214.3 to 245.6~GHz with a a spectral resolution of 0.6~MHz (0.78~km~s$^{-1}$). Observations of the CH$_3$OH $J_K$~=~5$_K-$4$_K$ branch at 241.791~GHz were made, and of $^{13}$CO $J$~=~2$-$1 at 220.398~GHz and of C$^{18}$O $J$~=~2$-$1 at 219.560~GHz were also made to compare methanol and CO emission. 

The data were calibrated and imaged using CASA\footnote{http://casa.nrao.edu/} \citep{McMullin2007}. The complex gains were calibrated through observations of the quasars 1751+096 and 1830+063, and the bandpass and the flux calibrations through observations of the quasar 3c279 and Callisto, respectively. The resulting SMA dataset has a typical beam-size of 7$\farcs$9~$\times$~4$\farcs$0 with a position angle of 68.8$^{\circ}$. 

Although the array was in this compact configuration, some emission is resolved out due to lack of short spacings. To map the extended emission towards the SVS~4 cluster, the SMA observations were therefore accompanied by short-spacing data obtained on May $10-14$, 2017, with the Atacama Pathfinder EXperiment (APEX; \citealt{Gusten2006}). The observations covered frequencies between 218.2~$-$~220.7~GHz and between 241.3~$-$~243.8~GHz, corresponding to the SMA 230~GHz receiver lower and 240~GHz receiver upper sidebands. 

\begin{table}[ht!]
\begin{center}
\caption{Sample of sources.}
\label{table:samples_sources}
\renewcommand{\arraystretch}{1.2}
\begin{tabular}{lccl} 
\hline \hline
Source &   RA    &  DEC  & $A_J$\\  
       & [J2000] & [J2000] & [mag]\\ \hline
SVS~4$-$2   & 18:29:56.58 & +01:12:59.4  & 4.2\\ 
SVS~4$-$3   & 18:29:56.70 & +01:12:39.0  & 18.0\\ 
SVS~4$-$4   & 18:29:57.00 & +01:12:47.6  & 4.9\\  
SVS~4$-$5   & 18:29:57.63 & +01:13:00.2  & 21.0\\ 
SVS~4$-$6   & 18:29:57.69 & +01:13:04.4  & 4.9\\  
SVS~4$-$7   & 18:29:57.84 & +01:12:27.8  & 10.5\\  
SVS~4$-$8   & 18:29:57.88 & +01:12:37.7  & 10.8\\  
SVS~4$-$9   & 18:29:57.92 & +01:12:46.0  & 10.5\\  
SVS~4$-$10  & 18:29:57.88 & +01:12:51.1  & 6.2\\  
SVS~4$-$12  & 18:29:57.22 & +01:12:58.4  & 26\\  
\hline
\end{tabular}
\end{center}
\footnotesize{\textbf{Notes.} The coordinates and the extinction in the $J$ band ($A_J$) are from \citet{Pontoppidan2004}. The estimated relative uncertainty on $A_J$ is 5\% \citep{Pontoppidan2004}.}
\end{table}

\begin{table*}
\begin{center}
\caption{Spectral data of the detected molecular transitions.}
\label{table:spectral_data_pointings}
\renewcommand{\arraystretch}{1.4}
\begin{tabular}{l c c l c c}
\hline \hline
 Transition & Frequency$^{a}$ & $A_\mathrm{ul}^{a}$ & $g_\mathrm{u}^{a}$ &  $E_\mathrm{u}^{a}$ & $n_\mathrm{cr}^{b}$ \\ 
            &  [GHz]    & [s$^{-1}$]& & [K] &  [cm$^{-3}$] \\ \hline 

C$^{18}$O $J$ = 2~$-$~1           & 219.560   &  6.01~$\times$~10$^{-7}$  &5     & 15.9  & 1.9~$\times~10^4$ \\ 
$^{13}$CO $J$ = 2~$-$~1           & 220.398   &  6.04~$\times~10^{-7}$    &5     & 15.9  & 2.0~$\times~10^4$ \\
CH$_3$OH $J$ = 5$_0~-~4_0$  E$^+$& 241.700   &  6.04~$\times~10^{-5}$    &11    & 47.9  & 5.5~$\times~10^5$ \\
CH$_3$OH $J$ = 5$_1~-~4_1$  E$^-$& 241.767   &  5.81~$\times~10^{-5}$    &11    & 40.4  & 4.8~$\times~10^5$ \\
CH$_3$OH $J$ = 5$_0~-~4_0$  A$^+$& 241.791   &  6.05~$\times~10^{-5}$    &11    & 34.8  & 5.0~$\times~10^5$ \\
CH$_3$OH $J$ =  5$_1~-~4_1$  E$^+$& 241.879   &  5.96~$\times~10^{-5}$    &11   & 55.9   & 4.6~$\times~10^5$ \\
CH$_3$OH $J$ = 5$_2~-~4_2$  E$^-$& 241.904   &  5.09~$\times~10^{-5}$    &11   & 60.7   & 4.2~$\times~10^5$ \\
\hline \hline
\end{tabular}
\end{center}

\begin{center}
\begin{tablenotes}
\small
\item{\textbf{Notes.}$^{a}$ From the Cologne Database for Molecular Spectroscopy (CDMS; \citet{Muller2001}) and the Jet Propulsion Laboratory catalog \citep{Pickett1998}. $^{b}$ Calculated using a collisional temperature of 10~K and collisional rates from the Leiden Atomic and Molecular Database (LAMDA; \citealt{Schoier2005}). The references for the collisional rates are \citet{Yang2010} for the CO isotopologues and \citet{Rabli2010} for CH$_3$OH.}
\end{tablenotes}
\end{center}
\end{table*}

The frequency resolution of the APEX observations is 0.076~MHz (0.099~km~s$^{-1}$). The map size achieved with APEX is 75$\arcsec$~$\times$~100$\arcsec$ , and it fully covers the region mapped by the SMA primary beams. The coordinates of the APEX pointing are $\alpha_{J2000} = 18^\mathrm{h}29^\mathrm{m}57^\mathrm{s}.76$, $\delta_{J2000} = +01^\circ 12\arcmin 49\farcs0$. A total integration time of 4.2 hours was adopted to match the sensitivity of the SMA data and to fold the datasets together. The APEX beam-size at the observed frequencies is 27$\farcs$4.
The reduction of the APEX dataset was performed with the GILDAS package CLASS\footnote{http://www.iram.fr/IRAMFR/GILDAS}. 
Subsequently, the reduced APEX datacube was imported to CASA to perform the combination of interferometric and single-dish data.

The SMA and APEX data were combined following the procedure described in Appendix~\ref{combination}. The species for which most of the flux is recovered in the combination is $^{13}$CO, followed by C$^{18}$O and CH$_3$OH. The interferometric observations missed up to 90\% of the emission, which implies that the gas column density would have been underestimated by approximately one order of magnitude. This illustrates how crucial the combination of interferometric and large-scale single-dish data is when the gas emission is extended and interferometric data have to be employed in a quantitative analysis (e.g. derivation of column densities).

\subsection{VLT observations} 
\label{VLT observations}
Archival data from the Very Large Telescope (VLT) were used to constrain the CH$_3$OH and CO ice content of the SVS-4 region. L- and M-band  spectra of ten SVS~4 sources (listed in Table \ref{table:samples_sources}) were obtained using the Infrared Spectrometer and Array Camera~(ISAAC) mounted on the Unit Telescope UT1-Antu of the VLT. All the L-band data have previously been published in \citet{Pontoppidan2004} and were observed as part of programme 71.C-0252(A). Briefly, a low-resolution grating and 0$\farcs$6 slit were used to cover the 2.80~--~4.15~$\mu$m spectral range, which yielded a resolving power of $\lambda/\Delta\lambda$~$\sim$~600. In the case of SVS~4$-$5 and SVS~4$-$9, the brightest sources, the 3.53~$\mu$m feature related to the methanol C-H stretching was observed with a resolving power of 3300. 

For the M-band observations, two sources (SVS~4$-$5 and SVS~4$-$9) were observed on May 6, 2002, as reported by \citet{Pontoppidan2003b}, and M-band spectra of eight other sources were obtained between May~2005 and August~2006, and are presented for the first time in this paper. The M-band spectra used the medium-resolution grating and the 0$\farcs$3 or 0$\farcs$6 slit, depending on source brightness, to yield resolving powers of $\lambda/\Delta\lambda$ 5~000~$-$~10~000 and a wavelength of 4.53~--~4.90~$\mu$m. Standard on-source integration times were $\sim$~45 min per pointing. The total integration time was doubled to 90 minutes for the faintest sources. The observations were carried out in pairs; each pointing was rotated to observe two stars at the same time. The data were reduced following the procedure described in \citet{Pontoppidan2003b}. The spectra were flux calibrated with respect to the standard stars HR 6629 and HR 7236 and wavelength calibrated to remove telluric absorption lines.   

\section{Results}
\label{results}
This section provides a summary of the methods we used to calculate the ice and gas-phase column densities presented in Sections~\ref{Ice_cd} $-$ \ref{Gas_cd}. Comprehensive descriptions of the methods applied when fitting the observational infrared data and  producing the gas-phase maps are given in Appendices~\ref{appendixA} and \ref{appendixB}.

\begin{figure*}
\centering
\includegraphics[trim={0 50 0 50},clip,width=6.6in]{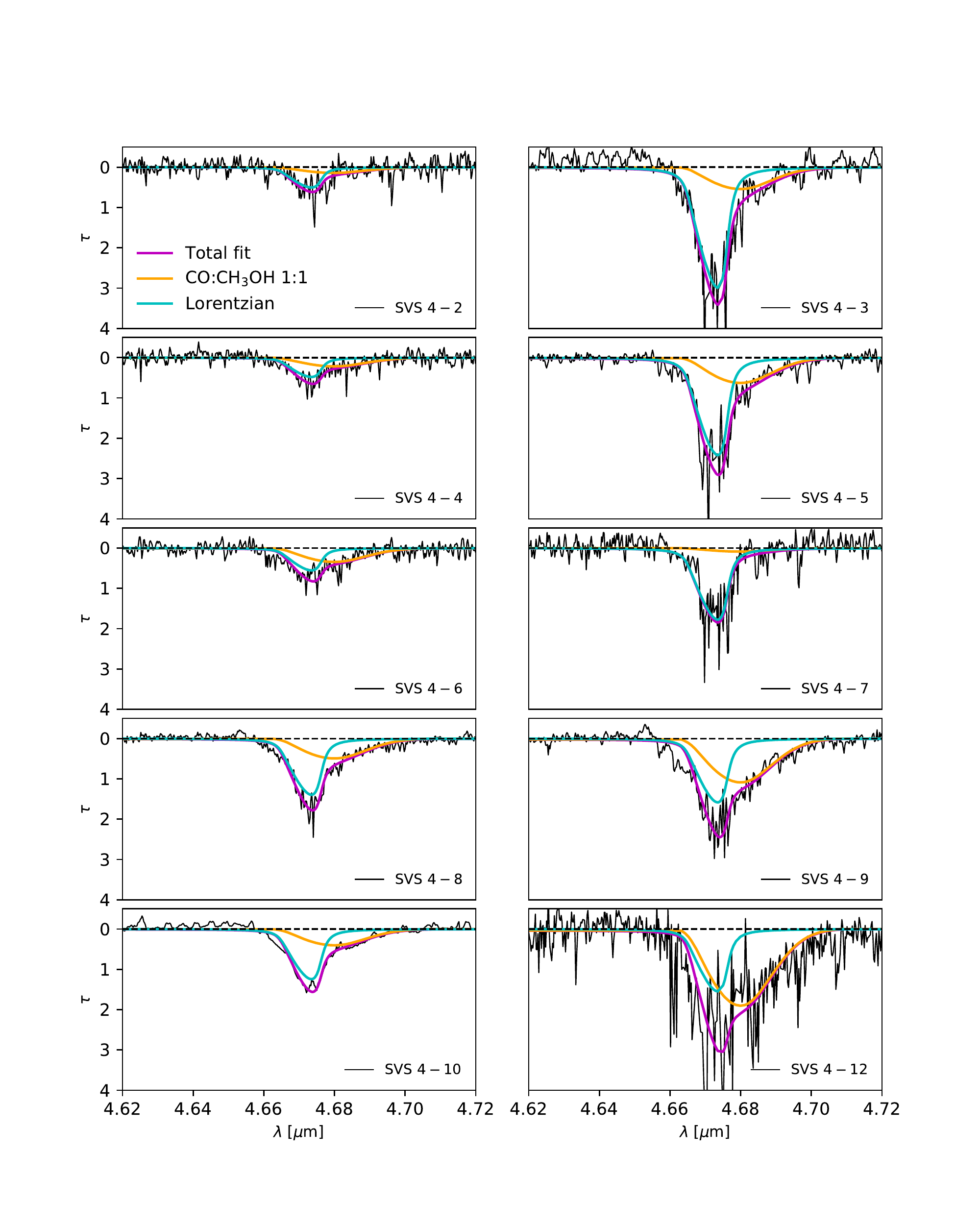}
\caption{M-band optical depth towards the SVS~4 sources (black). The coloured lines represent the CO:CH$_3$OH 1:1 laboratory spectrum (orange), the CDE-corrected Lorentzian reproducing the pure CO (cyan) and the total fit (magenta). The dashed black lines are used as a reference for $\tau$=0. The goodness of the fit $\chi_{\nu}^2$ starting from SVS~4$-$2 to SVS~4$-$12 is 0.95, 0.22, 0.79, 0.32, 0.83, 0.74, 0.30, 0.25, 0.87, and 0.64.}
\label{omnifitM}
\end{figure*}

\begin{figure*}
\centering
\includegraphics[trim={0 50 0 50},clip,width=6.5in]{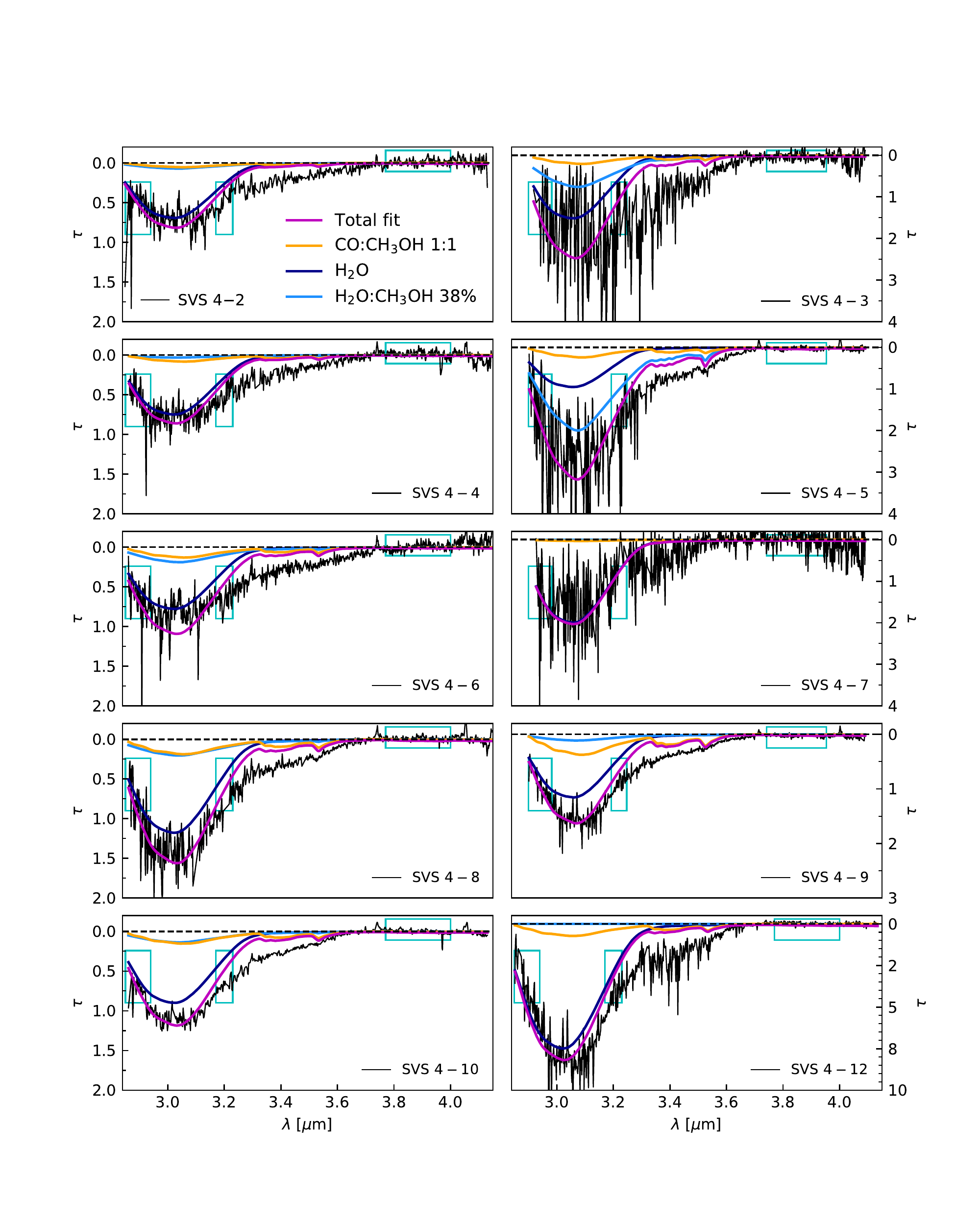}
\caption{L-band optical depth towards the SVS~4 sources (black). The coloured lines represent the laboratory ice data employed in the fitting: the pure H$_2$O (dark blue), the  H$_2$O:CH$_3$OH 38\% (light blue), the CO:CH$_3$OH 1:1 (orange), and the total fit (magenta). The dashed black lines are used as a reference for $\tau$~=~0. The cyan boxes illustrate the selected fitting ranges. The goodness of the fit $\chi_{\nu}^2$  starting from SVS~4$-$2 to SVS~4$-$12 is 0.59, 7.47, 0.55, 9.15, 0.79, 4.14, 0.72, 0.43, 0.25, and 22.5.}
\label{omnifitL}
\end{figure*}
\subsection{Ice column densities}
\label{Ice_cd}
To investigate the spatial distribution of ices towards the SVS~4 cluster, accurate ice column densities must be derived. To accomplish this, the spectrum was divided by a continuum and the optical depth was derived using the formula 
$\tau_{\lambda}~=~\mathrm{-ln}\ ({F_{\lambda}^{\mathrm{obs}}}/{F_{\lambda}^{\mathrm{cont}}}$), where $F_{\lambda}^{\mathrm{obs}}$ is the observed flux and $F_{\lambda}^{\mathrm{cont}}$ is the continuum.
A combination of blackbodies was used to determine the dust continuum for the L bands, whereas a spline function was adopted for the M bands. Table~\ref{table:photometry} in Appendix~\ref{appendixA} lists the photometric points used in the determination of the L-band continuum. 

Following the $\tau$ calculation, the spectral decomposition of the observational data was carried out with the OMNIFIT\footnote{https://ricemunk.github.io/omnifit/} fitting utility \citep{aleksi}: OMNIFIT is an open-source Python library used to fit observational data of astrophysical ices using multi-component laboratory spectra and analytical data. It makes use of the LMFIT package\footnote{https://lmfit.github.io/lmfit-py/} to perform the optimisation. In the M-band fitting, the non-linear optimisation by the Levenberg-Marquardt method was adopted, whereas the simplex Nelder-Mead and Monte Carlo Markov chain were used in the L-band fitting. The selected laboratory ice data were fitted to the observational spectra by applying a scale factor $k$. OMNIFIT provides a specific scale factor for each laboratory ice dataset.
Further details on the fitting utility can be found in Appendices~\ref{Optical depth fitting} and~\ref{Derivation of H$_2$O and CH$_3$OH ice column density}.

Table~\ref{table:omnifit_assumptions} lists the components and the fit parameters we adopted to fit the VLT-ISAAC data. OMNIFIT was applied in the same way to all the ten SVS~4 sources. 
To avoid fitting saturated features in the spectra, especially in the 3~$\mu$m region, only specific wavelength ranges were considered in the fits. For the L-band spectra, the 2.85~$-$~2.94~$\mu$m, 3.17~$-$~3.23~$\mu$m and 3.77~$-$~4.0~$\mu$m ranges were fitted. The first and second of these cover the blue wing and a portion of the red wing of the O-H stretching mode of H$_2$O ice, respectively. Finally, the third range was included because it was found to be a constraining factor to fit the C-H stretching mode of CH$_3$OH ice. It is worth noting that the typical C-H stretching region was not selected as a fitting range because of the effect of the 3.47~$\mu$m band and the scatter of the red wing of the O-H stretching mode. For the M bands, the whole spectral range, 4.62~$-$~4.72~$\mu$m, was fitted.

\begin{table}[hb!]
\caption{Fit parameters used to fit the infrared spectra with OMNIFIT.}
\label{table:omnifit_assumptions}
\centering
\setlength{\tabcolsep}{3pt} 
\renewcommand{\arraystretch}{1.4}
\scalebox{0.97}{
\begin{tabular}{c c c c}        
\hline\hline                 
\multicolumn{4}{c}{\bf $\textbf{L-band fitting}$}\\
Components & Fitting range~\tablefootmark{a}  &  Scale factor ($k$)~\tablefootmark{b} & Refs.~\tablefootmark{c} \\
\hline                   
pure H$_2$O      & 1 $-$ 3 & vary & 5\\
H$_2$O:CH$_3$OH  & 1 $-$ 3 & vary & 6\\
CO:CH$_3$OH      & 1 $-$ 3 & M-band values & 7 \\
\hline
\multicolumn{4}{c}{\bf $\textbf{M-band fitting}$}\\
Components   & Fitting range~\tablefootmark{a} & Scale factor ($k$)~\tablefootmark{b} & Refs.~\tablefootmark{c} \\
\hline
CO:CH$_3$OH & 4 & vary &  7 \\
CDE-Lorentzian & 4 & $-$ & 8 \\
\hline \hline
\end{tabular}
}
\begin{center}
\begin{tablenotes}
\small
\item{\textbf{Notes.}$^{a}$ The fitting ranges are [1] 2.85 $-$ 2.94~$\mu$m, [2] 3.17 $-$ 3.23~$\mu$m, [3] 3.77 $-$ 4.0~$\mu$m, and [4] 4.62 $-$ 4.72~$\mu$m. The goodness of the fits, $\chi_{\nu}^2$, was calculated for the specified fitting ranges. $^{b}$ Scale factor: multiplier of the spectrum in OMNIFIT (see Eq. \ref{A.2}.). $^{c}$ The references for the laboratory ice data are [5] \citet{Fraser2004}, [6] \citet{Dawes2016}, [7] \citet{Cuppen2011}, and [8] \citet{Pontoppidan2003b}}.
\end{tablenotes}
\end{center}
\end{table}

The spectral fitting procedure developed by \citet{Suutarinen2015} and adopted in this study stands as an alternative to the commonly used method to detect molecules in interstellar ices \citep{Grim1991,Brooke1999}. In contrast to the latter, which consists of fitting a polynomial to the red wing of the O-H stretching mode and analysing the residual optical depth at 3.53~$\mu$m to detect CH$_3$OH, the fitting procedure we adopted attempts to simultaneously fit multiple ice components to the 3~$\mu$m region to find the maximum amount of CH$_3$OH  that is hidden below the red wing. This approach makes it possible to concurrently account for both the C-H and the O-H stretches that are attributable to CH$_3$OH. This aspect is often overlooked, although the O-H stretching contribution of CH$_3$OH necessarily replaces some of the absorption of H$_2$O. The aim of this fitting procedure is to determine the maximum amount of CH$_3$OH ice that is hidden below the red wing of interstellar ices, and thus to help explain the high CH$_3$OH abundances that are observed in the gas-phase.

Figure~\ref{omnifitM} shows the optical depth of the sources listed in Table~\ref{table:samples_sources} decomposed by analytical and experimental components using the OMNIFIT fitting utility \citep{aleksi}. As outlined in \citet{Pontoppidan2003b}, the M-band features can be decomposed by a linear combination of three components: blue, middle, and red. We address only the middle and red components because the blue component does not constrain the methanol ice abundance, but rather that of CO$_2$ ice \citep{Boogert2015}. The central peak (i.e. middle component) is associated with the C-O vibrational mode of pure CO ice. It was fitted using a continuous distribution of ellipsoids (CDE) corrected Lorentzian function in order to take into account irregularities in the grain shape and match the central peak of the observations \citep{Pontoppidan2003b}.

The red component, on the other hand, was fitted with CO:CH$_3$OH laboratory data \citep{Cuppen2011} because the observed CO peak position is slightly shifted to longer wavelengths, possibly due to migration of CO in the ices observed towards the cluster sources \citep{Devlin1992,Zamirri2018}. Although the red component can be associated with CO:H$_2$O ice \citep{Thi2006}, the CO:CH$_3$OH mixture is the most likely carrier of this feature \citep{Cuppen2011}, and it is assumed to trace warmer regions than pure CO. The goodness of the fit was addressed using the reduced $\chi_{\nu}^2$, that is given by $\chi_{\nu}^2 = \chi^2/\nu$, where $\nu$ is the difference between the number of data points and the number of free parameters in the fit.

The decomposition of the L-band features, between 2.80~$-$~4.15~$\mu$m, was carried out using three experimental components shown in Figure~\ref{omnifitL}: pure H$_2$O in blue \citep{Fraser2004}, H$_2$O:CH$_3$OH in cyan \citep{Dawes2016}, and CO:CH$_3$OH in orange \citep{Cuppen2011}. Pure H$_2$O was selected because it is the most abundant ice component. H$_2$O:CH$_3$OH was chosen because CH$_3$OH is the second major contributor to the O-H stretching at 3~$\mu$m, after H$_2$O. H$_2$O:CH$_3$OH was preferred over pure CH$_3$OH as the C$-$H stretching of the latter is too prominent and shifted in the observational data. As \citet{Dawes2016} showed, the O-H spectroscopic profile and exact peak position of the H$_2$O:CH$_3$OH mixture are dependent not only on the host molecule of the mixture (H$_2$O), but also on the minor component (CH$_3$OH). CH$_3$OH is thought to form on dust grains from hydrogenation of CO ;this may occur early, that is, during the early stages of CH$_3$OH formation, and involves reactions of HCO + OH, or it may occur late after critical CO freeze-out. Ice studies of binary and tertiary systems show that even if ices form in layers, the heating and processing with time and diffusion will lead to ice mixing. In addition, H$_2$O:CH$_3$OH was used because recent modelling and laboratory experiments showed evidence of the production of CH$_3$OH mixed in H$_2$O matrices upon UV and X-ray irradiation of simple ice components, for instance, H$_2$O:CO \citep{Jimenez-Escobar2016, MunozCaro2019} or by interaction of CH$_4$, O$_2$ and H \citep{Qasim2018}. At the same time, mixtures such as NH$_3$:H$_2$O or CO$_2$:H$_2$O were discarded because they would have very little effect on the shape and spectroscopy of the O-H stretching at 3~$\mu$m and they would reduce the contribution of CH$_3$OH, which would prevent us from determining the maximum amount of methanol that is hidden below the red wing. Finally, CO:CH$_3$OH was selected to ensure that the fit of the methanol band at 3.53~$\mu$m is consistent with the fit of the CO red component of the M-band spectra. To do so, the $k$ values for CO:CH$_3$OH obtained in the M-band fitting were carried over in the L-band fitting. For all the other laboratory ice data, $k$ varies. The same approach was also used by \citet{Suutarinen2015} to fit the L-band features of several YSOs, including SVS~4$-$5 and SVS~4$-$9. It is worth noting that the 3~$\mu$m water band of SVS~4$-$12 is saturated (see Figure \ref{photometry_plot_Lp2}), and therefore, the optical depth below $\lambda = 3.3~\mu$m for this source was determined by scaling the SVS 4$-$9 L-band features to the SVS 4$-$12 red wing following the procedure presented in \citet{Pontoppidan2004}. The best scaling was found for $\tau_\mathrm{max}^\mathrm{L-band}~=~9$. 

Despite the degeneracies involved in the spectral fitting, Figure~3 shows that in addition to the contribution of the O-H stretching of methanol, the pure H$_2$O component mainly reproduces the 3~$\mu$m absorption band. This is in agreement with a scenario where the predominant ice constituent is pure water, but it also supports the hypothesis of a degree of mixing in astrophysical ices. For most of the spectra, a third component, the H$_2$O:CH$_3$OH mixture, is required to obtain a better fit constrained by the adopted fitting ranges.
For some sources, for example SVS~4$-$2, SVS~4$-$4, and SVS~4$-$6, the selection of the fitting ranges results in an overfit of the O-H stretching mode. Given the low signal-to-noise ratio (S/N) of the observations, this study does not aim to prove that H$_2$O:CH$_3$OH models the observational spectra better than other laboratory data. It was selected to find the maximum amount of CH$_3$OH that is hidden in the H$_2$O ice, if the L band is decomposed with the laboratory data selected in this study. 

It is worth noting that the CH$_3$OH band at 3.53~$\mu$m is not constrained by the fitting procedure itself, but rather by the methanol fraction in both CO:CH$_3$OH and H$_2$O:CH$_3$OH mixtures. The amount of CH$_3$OH estimated from the M-band fitting results in sufficient absorption to model the C$-$H stretching mode at 3.53~$\mu$m, and therefore the addition of H$_2$O:CH$_3$OH, aimed to fit the constrained ranges, has a minor contribution in the 3.3$-$3.6~$\mu$m interval. The only two exceptions are SVS~4$-$3 and SVS~4$-$5 for which a greater contribution of the H$_2$O:CH$_3$OH mixture is observed. Overall, the fit of the C$-$H stretching mode at 3.53~$\mu$m is satisfactory because the model is predicting an absorption lower than or equal to the observed band, as well as taking into account that other molecules and physical processes, such as scattering by large grains, contribute to this spectral range. Reproducing the full range of absorptions observed in the red wing was not part of the scope of this study, and the selected fitting ranges would indeed prevent a good quality fit of it.
 
\begin{figure}
\centering
\includegraphics[trim={5 95 50 160},clip,width=3.55in]{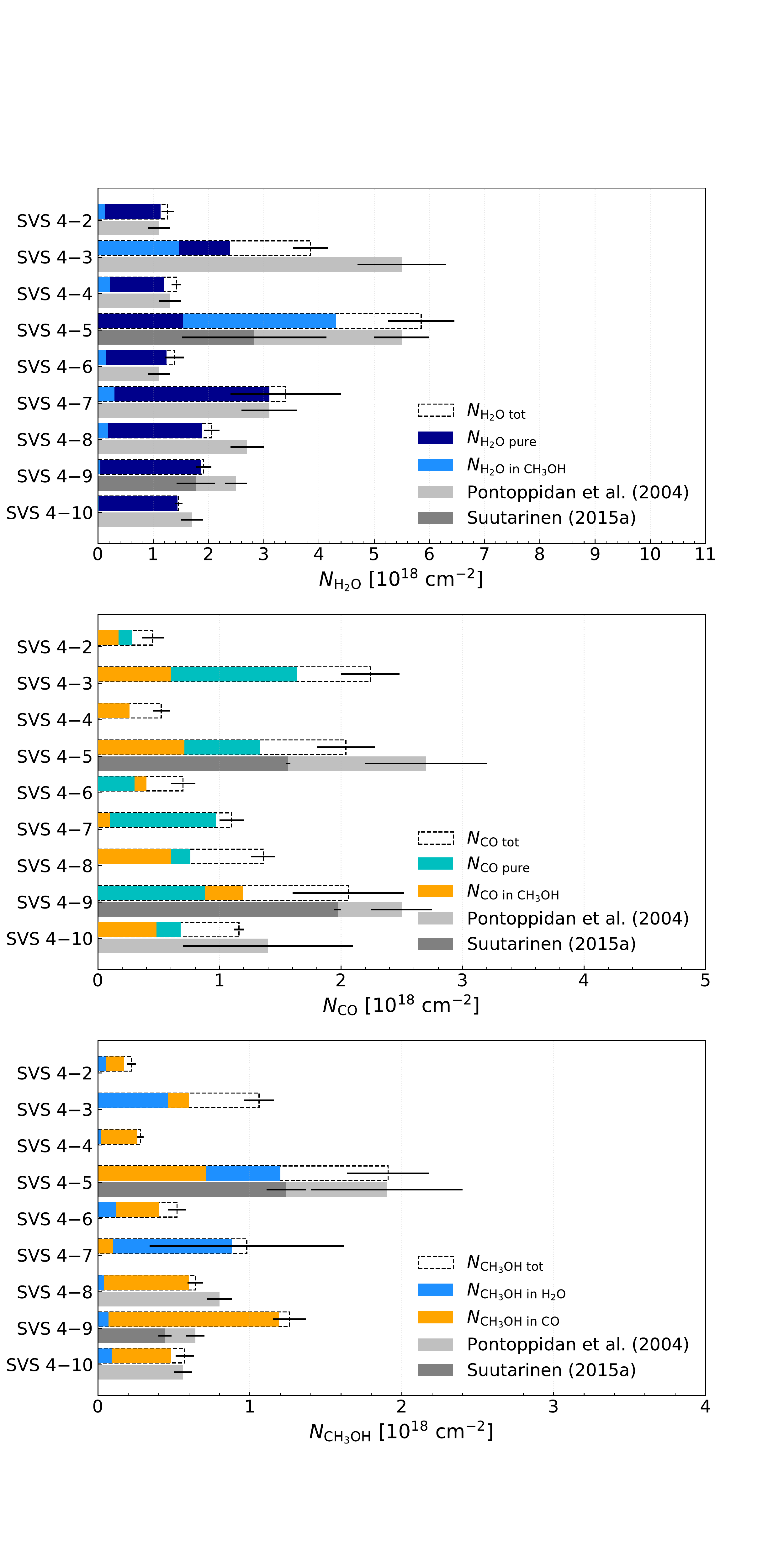}
\caption{H$_2$O (top), CO (middle), and CH$_3$OH (bottom) ice column densities obtained in this study compared to \citet{Pontoppidan2004} (light grey) and \citet{Suutarinen2015} (dark grey). The CO column density towards SVS~4$-$10 is from \citet{Chiar1994}. The coloured bars represent the ice column densities of the pure species or of the species in a mixture, and the dashed bars depict the total ice column densities obtained by summing the ice column densities of the pure species and of the species in a mixture.}
\label{barplot}
\end{figure}

\begin{table*}
\begin{center}
\caption{Total ice and gas column densities of the species analysed towards the SVS~4 sources.}
\label{table:summary_cd}
\renewcommand{\arraystretch}{1.8}
\begin{tabular}{llllllll} 
\hline \hline
Source & $N^\mathrm{ice}_\mathrm{H_2O\ tot}$ & $N^\mathrm{ice}_\mathrm{CO \ tot}$ & $N^\mathrm{ice}_\mathrm{CH_3OH \ tot}$ & $N^\mathrm{gas}_\mathrm{CO \ tot}$ & $N^\mathrm{gas}_\mathrm{CH_3OH \ tot}$  & $N_\mathrm{H_2}^\mathrm{SCUBA-2}$ & $^{*}N_\mathrm{H_2}^{A_J}$ \\  
      &[10$^{18}$cm$^{-2}$]&  [10$^{18}$cm$^{-2}$] &[10$^{18}$cm$^{-2}$] & [10$^{18}$cm$^{-2}$] & [10$^{14}$cm$^{-2}$] & [10$^{23}$cm$^{-2}$]    & [10$^{23}$cm$^{-2}$] \\ \hline 
SVS~4$-$2   & 1.26 $\pm$ 0.11 & 0.45 $\pm$ 0.08 & 0.22 $\pm$ 0.03 & 2.71 $^{+0.55}_{-0.13}$ & 8.24 $^{+12.6}_{-1.70}$ & 0.80 $^{+0.90}_{-0.30}$  & 0.14 \\  
SVS~4$-$3   & 3.85 $\pm$ 0.32 & 2.24 $\pm$ 0.20 & 1.06 $\pm$ 0.10 & 2.58 $^{+0.53}_{-0.12}$ & 1.92 $^{+1.53}_{-0.10}$  & 0.18 $^{+0.20}_{-0.06}$ & 0.6      \\  
SVS~4$-$4   & 1.42 $\pm$ 0.09 & 0.52 $\pm$ 0.05 & 0.28 $\pm$ 0.02 & 2.65 $^{+0.53}_{-0.13}$ & 3.23 $^{+2.55}_{-0.13}$  & 0.30 $^{+0.34}_{-0.11}$ & 0.16 \\  
SVS~4$-$5   & 5.85 $\pm$ 0.60 & 2.04 $\pm$ 0.19 & 1.91 $\pm$ 0.27 & 3.70 $^{+0.75}_{-0.18}$ & 4.50 $^{+6.80}_{-0.87}$  & 0.65 $^{+0.73}_{-0.24}$ & 0.7      \\  
SVS~4$-$6   & 1.38 $\pm$ 0.17 & 0.70 $\pm$ 0.08 & 0.52 $\pm$ 0.06 & 3.20 $^{+0.65}_{-0.16}$ & 6.67 $^{+10.2}_{-1.29}$ & 0.78 $^{+0.87}_{-0.28}$  & 0.16 \\  
SVS~4$-$7   & 3.40 $\pm$ 1.00 & 1.10 $\pm$ 0.09 & 0.98 $\pm$ 0.64 & 2.80 $^{+0.57}_{-0.14}$ & 3.27 $^{+4.98}_{-0.59}$  & 0.16 $^{+0.18}_{-0.06}$ & 0.35 \\  
SVS~4$-$8   & 2.06 $\pm$ 0.14 & 1.36 $\pm$ 0.07 & 0.64 $\pm$ 0.05 & 2.77 $^{+0.56}_{-0.14}$ & 3.10 $^{+2.44}_{-0.14}$  & 0.20 $^{+0.22}_{-0.07}$ & 0.36 \\  
SVS~4$-$9   & 1.91 $\pm$ 0.14 & 2.06 $\pm$ 0.37 & 1.26 $\pm$ 0.11 & 2.35 $^{+0.48}_{-0.11}$ & 2.17 $^{+2.72}_{-0.32}$  & 0.28 $^{+0.32}_{-0.10}$ & 0.35 \\  
SVS~4$-$10  & 1.46 $\pm$ 0.07 & 1.16 $\pm$ 0.03 & 0.57 $\pm$ 0.06 & 2.66 $^{+0.54}_{-0.13}$ & 2.81 $^{+2.81}_{-0.27}$  & 0.38 $^{+0.42}_{-0.14}$ & 0.2      \\  
SVS~4$-$12 & 12.4 $\pm$ 5.90 & 3.12 $\pm$ 0.74 & 2.28 $\pm$ 0.18 & 3.32 $^{+0.68}_{-0.16}$ & 3.14 $^{+4.77}_{-0.60}$  & 0.73 $^{+0.82}_{-0.26}$  & 0.86 \\  
\hline 
\end{tabular}
\end{center}
\footnotesize{\textbf{Notes.} Columns 5$-$7 list the column densities that were calculated for $T$~=~15~K. For $N^\mathrm{gas}_\mathrm{CO}$, $N^\mathrm{gas}_\mathrm{CH_3OH}$ , and $N_\mathrm{H_2}$, the values in the superscript and subscript represent the upper and lower limits if the column densities are calculated at $T$~=~10~K or $T$~=~20~K, respectively. $^{*}$ Estimated relative uncertainty of 5\%. See Sections~\ref{Ice_cd}$-$\ref{Gas_cd} and Appendices~\ref{appendixA}, \ref{appendixB}, and \ref{appendixC} for a more detailed description of the determination of ice, gas, and H$_2$ column densities.}
\end{table*}

\begin{table}
\begin{center}
\caption{Ice column densities of the pure and mixed components given as percentages of the total column densities.}
\label{table:summary_cd_ice}
\renewcommand{\arraystretch}{1.7}
\scalebox{0.98}{
\begin{tabular}{lccc} 
\hline \hline
Source & $\frac{N^\mathrm{ice}_\mathrm{H_2O\ pure}}{N^\mathrm{ice}_\mathrm{H_2O\ tot}}$ \footnotesize{[\%]} & $\frac{N^\mathrm{ice}_\mathrm{CO \ pure}}{N^\mathrm{ice}_\mathrm{CO \ tot}}$ \footnotesize{[\%]} & $\frac{N^\mathrm{ice}_\mathrm{CH_3OH\ in\ H_2O}}{N^\mathrm{ice}_\mathrm{CH_3OH \ tot}}$ \footnotesize{[\%]} \\ 
\hline 
SVS~4$-$2   & 89.7 $\pm$ 11.7 & 62.2 $\pm$ 21.7 & 22.7 $\pm$ 13.9 \\  
SVS~4$-$3   & 62.1 $\pm$ 8.50 & 73.2 $\pm$ 11.6 & 43.4 $\pm$ 9.43 \\  
SVS~4$-$4   & 84.5 $\pm$ 8.30 & 50.0 $\pm$ 11.7 & 7.10 $\pm$ 3.61 \\  
SVS~4$-$5   & 26.3 $\pm$ 6.56 & 65.2 $\pm$ 11.7 & 62.8 $\pm$ 16.3 \\  
SVS~4$-$6   & 89.9 $\pm$ 16.0 & 42.9 $\pm$ 11.7 & 23.1 $\pm$ 9.98 \\  
SVS~4$-$7   & 91.2 $\pm$ 39.8 & 88.2 $\pm$ 11.5 & 89.8 $\pm$ 87.8 \\  
SVS~4$-$8   & 91.3 $\pm$ 8.84 & 55.9 $\pm$ 5.51 & 6.20 $\pm$ 1.64 \\  
SVS~4$-$9   & 97.9 $\pm$ 10.3 & 42.2 $\pm$ 20.7 & 5.6  $\pm$ 3.21 \\  
SVS~4$-$10  & 98.6 $\pm$ 6.73 & 58.6 $\pm$ 2.65 & 15.8 $\pm$ 10.7 \\  
SVS~4$-$12  & 97.7 $\pm$ 66.3 & 27.3 $\pm$ 23.9 & 0.40 $\pm$ 0.44 \\  
\hline 
\end{tabular}
}
\end{center}
\footnotesize{\textbf{Notes.} To avoid redundancy, the percentages of ice column densities of H$_2$O in CH$_3$OH, CO in CH$_3$OH, and CH$_3$OH in CO w.r.t. the total H$_2$O, CO, and CH$_3$OH ice column densities are not tabulated. They are equal to 100 minus the value reported in the corresponding column (e.g. 10.3 cm$^{-2}$ for H$_2$O in CH$_3$OH towards SVS~4$-$2).} See Sections~\ref{Ice_cd} and Appendix~\ref{appendixA} for a more detailed description of the determination of ice column densities.
\end{table}

The CO, H$_2$O, and CH$_3$OH ice column densities were obtained from the integrated $\tau$ of the components found by the fitting routine. The calculation of the ice column densities is described in detail in Appendices \ref{Derivation of H$_2$O and CH$_3$OH ice column density}$-$\ref{Derivation of CO ice column density} and their values are shown in Tables~\ref{table:summary_cd} and \ref{table:summary_cd_ice}. In these tables, the total ice column density is the sum of the pure and mixed components. SVS~4$-$12 shows the deepest absorptions features, thus the column densities are the highest towards this source. The uncertainties in the column density were calculated from the error in the optical depth.

Figure~\ref{barplot} compares the derived ice column densities between the SVS~4 sources (Table~\ref{table:summary_cd}) to measurements from the literature  \citep{Pontoppidan2004, Suutarinen2015}. Because of the low S/N, SVS~4$-$12 is not included in this figure. In the case of H$_2$O ice, shown in the upper panel, the pure water component is the most abundant compared to the mixed water, except in the case of SVS~4$-$7. This result indicates a negligible level of molecular migration in the ice matrix. It is also worth observing that the H$_2$O ice column density is diminished by the contribution of the O-H stretching associated with CH$_3$OH. In the case of CO ice, the pure CO column density dominates the mixed component, except in the case of SVS~4$-$4, SVS~4$-$6 and SVS~4$-$9, which suggests that the circumstellar environment around these three objects is warmer than in the other regions in the SVS~4 cluster. In one case, SVS~4$-$4, the column densities of pure CO and CO in the CO:CH$_3$OH mixture are the same.

\begin{figure*}
\centering
\includegraphics[trim={0 0 0 0},clip,width=6in]{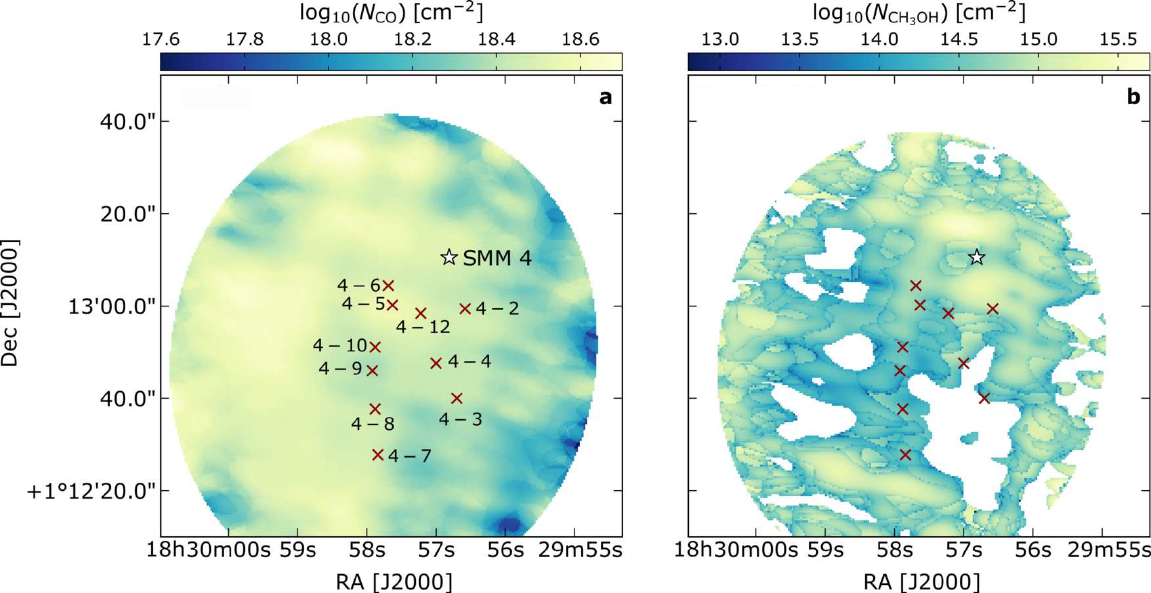}
\caption{Gas column density maps of the SVS~4 cluster. \textit{a:} $^{12}$CO column density map. \textit{b:} CH$_3$OH column density map. The white spaces within the area covered by SMA+APEX are regions where the strength of the emission drops below 5$\sigma$. The red crosses mark the position of the SVS~4 sources, and the white star represents SMM~4.}
\label{discussion_plots}
\end{figure*} 

For CH$_3$OH, the CO:CH$_3$OH component dominates in all the sources, except for SVS~4$-$5 and SVS~4$-$7. This indicates that the amount of methanol constrained from the M-band fitting is sufficient to reproduce the methanol absorption in the L band. SVS~4$-$4 shows the same behaviour as observed in the CO case because the column densities of CH$_3$OH in CO:CH$_3$OH and H$_2$O:CH$_3$OH are the same. When the values estimated in this study can be compared to the literature, the total ice methanol column densities agree well with those reported by \citet{Pontoppidan2004}, except for SVS~4$-$9 because of uncertainties in the local continuum determination for this particular source. SVS~4$-$5 has the highest methanol column density in both studies except for SVS~4$-$12. The column densities given by \citet{Suutarinen2015} are generally a factor of $\sim$1$-$3 lower.  
The average total ice column densities of H$_2$O, CO, and CH$_3$OH are $\mathrm{2.5\times10^{18} \; cm^{-2}}$, $\mathrm{1.3\times10^{18} \; cm^{-2}}$ , and $\mathrm{0.8\times10^{18} \; cm^{-2}}$, respectively. The CO and CH$_3$OH average abundances with respect to H$_2$O are 0.52 and 0.32. The high CO fraction relative to H$_2$O indicates an environment for the SVS~4 cluster with a temperature below the CO desorption ($\sim$20~K), and it agrees with  the average ice column densities of low-mass stars shown in \citet{Oberg2011_spitzer}. On the other hand, the CH$_3$OH fraction with respect to H$_2$O ice is a factor of 2.6 higher than the upper quartile values calculated in \citet{Oberg2011_spitzer}. Nevertheless, this value is consistent with the high methanol ice abundance for the SVS~4 cluster measured by \citet{Pontoppidan2004}, namely 0.28, which suggests a peculiar efficiency of methanol ice synthesis in this region of the Serpens molecular core.

\subsection{Gas column densities}
\label{Gas_cd}
The CO isotopologue and CH$_3$OH gas column densities were calculated from the integrated line intensities from the combined SMA~+~APEX maps. Table~\ref{table:spectral_data_pointings} lists the spectral line data of the detected molecular transitions. All the gas column densities derived in this section and their uncertainties are listed in Table~\ref{table:summary_cd}, and the CO and CH$_3$OH column density maps are displayed in Figure~\ref{discussion_plots}.
Towards some of the SVS~4 sources, the $^{13}$CO emission is found to be moderately optically thick (see Appendix~\ref{line_ratio}). As a result, the $^{13}$CO column densities are underestimated towards these positions. Therefore the optically thin C$^{18}$O emission is used to estimate the column density of the CO gas. The C$^{18}$O column densities are obtained using the combined SMA~+~APEX integrated intensity of the C$^{18}$O $J = $~2$-$1 line at 219.560~GHz, assuming a kinetic temperature equal to 15~K and optically thin emission. The derived C$^{18}$O column densities are of the order of 10$^{15}$~cm$^{-2}$ towards all the SVS~4 star positions. The C$^{18}$O column densities are then converted into $^{12}$CO column densities using a $^{16}$O/$^{18}$O isotope ratio of 557~$\pm$~30 \citep{Wilson1999}. The calculated $^{12}$CO column densities are of the order of  10$^{18}$~cm$^{-2}$ towards the SVS~4 stars (Figure~\ref{discussion_plots}). 

The gas column densities of CH$_3$OH towards the SVS~4 sources were calculated from the combined SMA~+~APEX integrated intensities by assuming local thermodynamic equilibrium (LTE) and that the emission of methanol is optically thin. A kinetic temperature equal to 15~K was selected in this case as well, representing the global temperature of the cluster (see Section \ref{H2_cd}). The measured CH$_3$OH column densities are of the order of 10$^{14}$~cm$^{-2}$ towards all the SVS~4 sources (Figure~\ref{discussion_plots}). Similar column densities are also reported in \citet{Kristensen2010}, especially towards two outflow knot positions nearby SMM~4 labelled SMM~4$-$S and SMM~4$-$W and in \citet{Oberg2009a} towards SVS~4$-$5. In case of sub-thermal excitation the methanol column density would be overestimated by up to one order of magnitude \citep{Bachiller1995,Bachiller1998}.


\section{Analysis}
\label{analysis}

\subsection{H$_2$ column density and physical structure of the region}
\label{H2_cd}

The SVS~4 sources might be embedded within the envelope of SMM~4 to different degrees and not be uniformly located behind it \citep{Pontoppidan2004}. As a result, the absolute ice column densities might not be the most appropriate measure to compare directly to gas-phase column densities. The gas and ice observations might be tracing different columns of material. To address this point, a search for an ice-gas correlation was conducted by comparing gas and ice abundances relative to the H$_2$ column density of the region.
The abundances of the gas species were calculated by dividing the gas column densities by the H$_2$ column density estimated from the submillimeter continuum (SCUBA-2) map at 850~$\mu$m \citep{Herczeg2017}, assuming that the continuum radiation comes from optically thin thermal dust emission (Appendix~\ref{appendixC} and \citealt{Kauffmann2008}). Under these assumptions, the strength of the submillimeter continuum emission depends on the dust temperature, $T$, on the column density, and on the opacity, $\kappa_{\nu}$. 

The dust temperature for the SVS~4 region has previously been estimated to lie in the range of 10~K~$\leq T \leq$~20~K according to \citet{Schnee2005} and to the dust radiative transfer model presented in \citet{Kristensen2010}. The H$_2$ column density is thus calculated for $T$~=~10~K, 15~K and 20~K. We adopted the opacities corresponding to dust grains covered by a thin ice mantle \citep{Ossenkopf1994} with $\kappa_{\nu}$ equal to 0.0182 cm$^2$ g$^{-1}$ at 850~$\mu$m ("OH5 dust"). When a temperature of 15~K is assumed, all the sources lie in the region where the H$_2$ column density is of the order of 10$^{22}$~cm$^{-2}$ (see Figure~\ref{NH2_map} and Table~\ref{table:summary_cd}). The estimated gas abundances relative to H$_2$ range from $10^{-5}-10^{-4}$ for CO and $10^{-9}-10^{-8}$ for CH$_3$OH at the SVS~4 source positions. The latter agree with the values measured by \citet{Kristensen2010} towards the Serpens molecular core.

The H$_2$ column densities derived from the SCUBA-2 measurements using the method above provide a good estimate of the total beam-averaged amount of gas and are therefore useful as a reference for the optically thin gas-phase tracers. However, it is not clear that they can directly be related to the ice data that provide (pencil-beam) measurements of the column densities towards the infrared sources that may or may not be embedded within the cloud. For ice abundances we therefore adopted a slightly different method of deriving the H$_2$ column densities, starting from extinction in the $J$-band $A_J$ (Table~\ref{table:samples_sources}) and adopting the relation $N_\mathrm{H_2}$=3.33~$\times$~10$^{21}A_J$ given in Figure~10 of \citet{Pontoppidan2004}. At the SVS~4 source positions, the H$_2$ column density is of the order of 10$^{22}$~cm$^{-2}$ and the measured ice abundances relative to H$_2$ are of the order of $10^{-5}$ for CO and CH$_3$OH and $10^{-5}-10^{-4}$ for H$_2$O ice. The ice abundances agree with the values reported by \citet{Pontoppidan2004} for the SVS~4 cluster. The column densities derived using the two different methods are of the same order of magnitude, but with small variations, possibly due to the assumptions about the temperature, the exact location of the infrared sources along the LoS compared to the distribution of the cloud material, and the differences in the exact column densities traced by the SCUBA-2 and $A_J$ measurements.

\begin{figure}
\centering
\includegraphics[trim={0 0 0 0},clip,width=3.in]{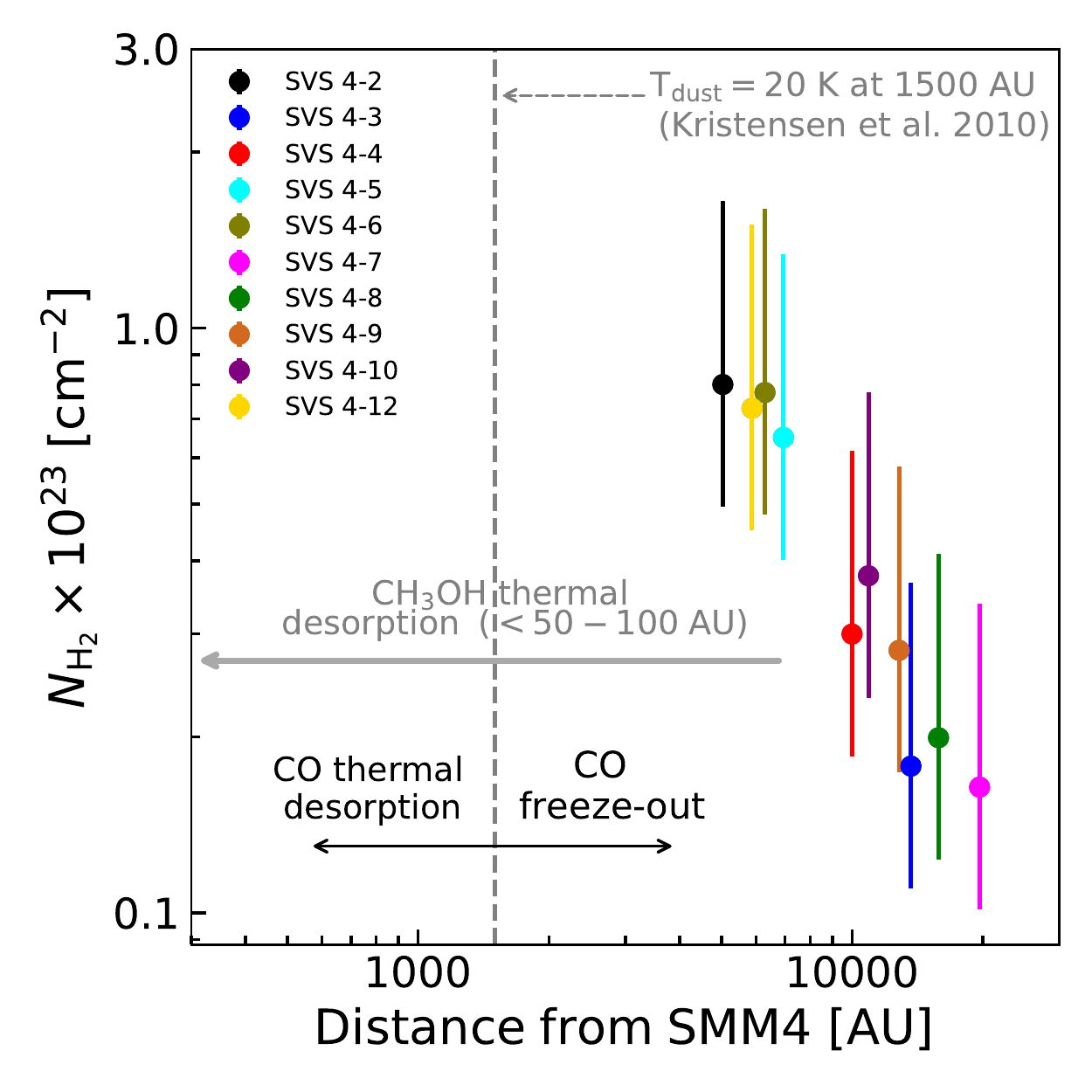}
\caption{H$_2$ column density estimated from the 850 $\mu$m dust emission map \citep{Herczeg2017} as a function of the distance from SMM~4. A distance of 433~pc is assumed \citep{Ortiz_Leon2018} in order to show the distance in astronomical units. All the sources are located in both CO and CH$_3$OH freeze-out zones, as indicated by the solid arrows and the grey dashed line, as taken from the DUSTY model in \citet{Kristensen2010}.}
\label{scuba_vs_aj}
\end{figure}

Figure~\ref{scuba_vs_aj} shows the $N_\mathrm{H_2}$ variation as a function of the distance from the SMM~4 position for each SVS~4 source. The upper and lower error bars in this figure represent the H$_2$ column density calculated for $T$~=~10~K and 20~K, respectively. The vertical dashed line indicates the dust temperature of 20~K at 1500~AU from SMM~4, calculated from \citet{Kristensen2010} by scaling the luminosity of the protostar according to the most updated distance measurements obtained with $Gaia$-$DR2$ (433~pc~$\pm$~4; \citealt{Ortiz_Leon2018}). The DUSTY model assumes a spherically symmetric envelope and that the central heating source is a blackbody at $T$~=~5000~K. The black rightward arrow indicates the distance where CO gas is expected to freeze out on dust grains, which enhances the CH$_3$OH formation through CO hydrogenation. Pure CO ice starts to desorb at $T$~>~20~K (black leftward arrow), although some frozen CO molecules migrate to porous water ice and are only desorbed at $T$~>~90~K \citep{Collings2003}. According to the model, the methanol ice desorption region is located below 50~$-$~100~AU, as indicated by the solid grey arrow in the figure. The comparison with the model shows that the SVS~4 sources lie in a region in which both CO and CH$_3$OH are frozen out. The observed gas emission associated with these molecules, however, has been released to the gas-phase by non-thermal desorption mechanisms.

\begin{figure}
\includegraphics[trim={135 70 70 7},clip,width=3.2in]{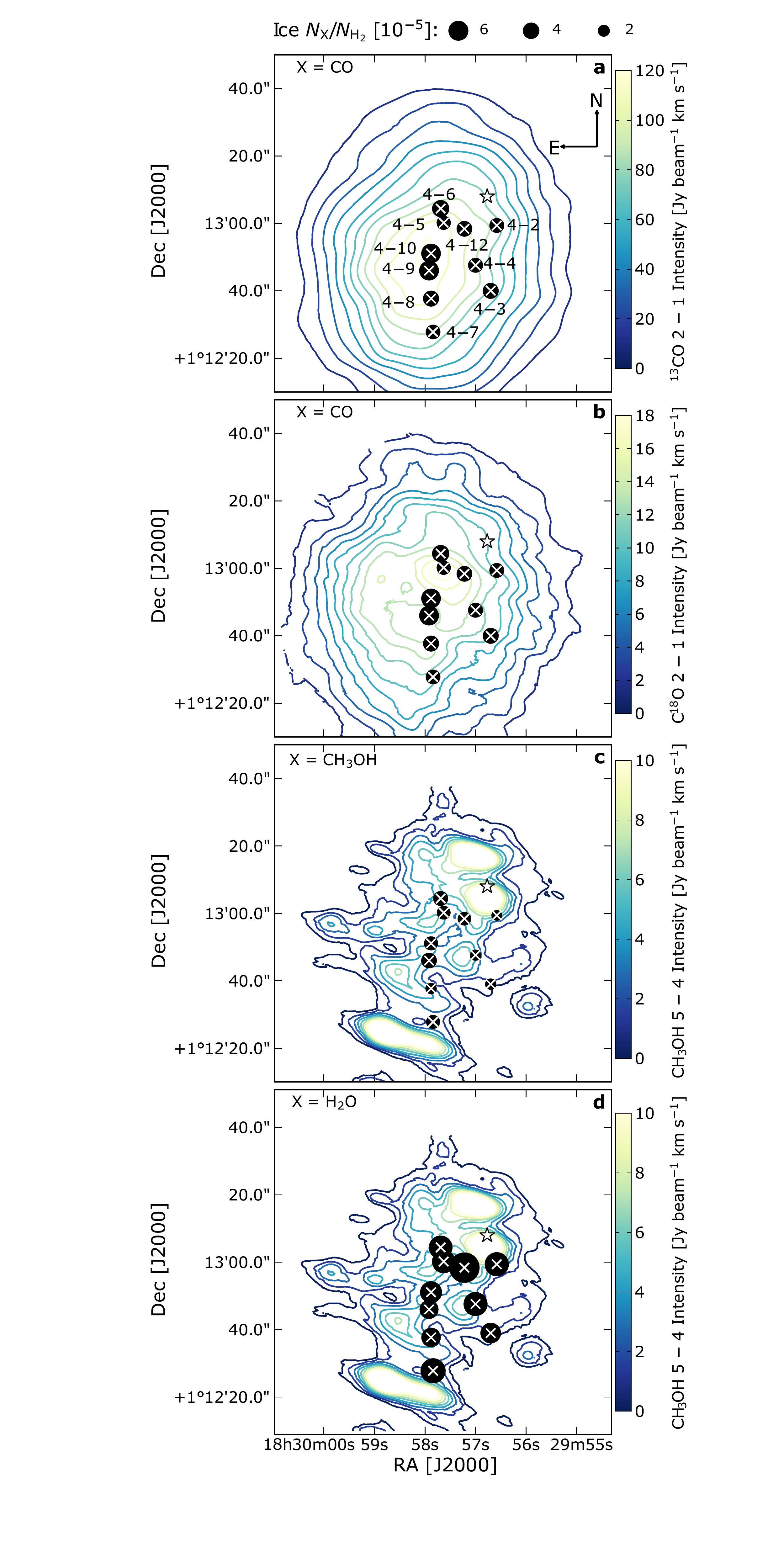}
\caption{Gas-ice maps of the SVS~4 cluster. Ice abundances are shown as black circles. The contours at 3$\sigma$, 6$\sigma$, 9$\sigma$, etc. represent the gas-phase integrated intensities. \textit{a:} CO abundances on gas $^{13}$CO~2~$-$~1 ; \textit{b:} CO abundances on gas C$^{18}$O~2~$-$~1 ; \textit{c:} CH$_3$OH abundances on gas CH$_3$OH~5$_0~-~4_0$~A$^+$. \textit{d:} H$_2$O abundances
on gas CH$_3$OH~5$_0~-~4_0$~A$^+$. The white crosses mark the position of the SVS~4 sources, and the white star represents SMM~4.}
\label{ig_maps}
\end{figure}

\begin{figure*}
\centering
\includegraphics[trim={0 10 0 30},clip,width=7.3in]{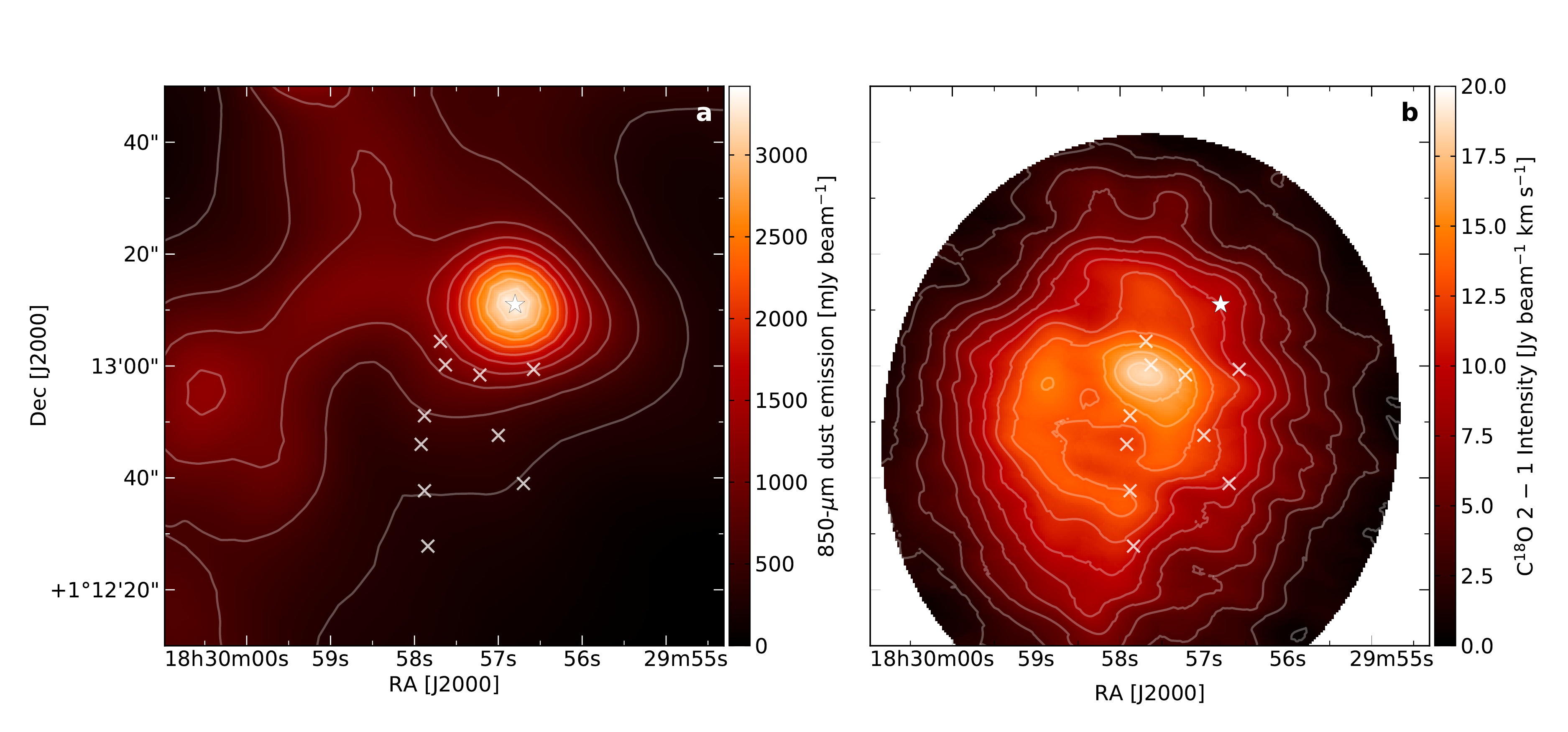}
\caption{$a$: SCUBA-2 850~$\mu$m dust continuum emission (contours decrease in steps of 10\% starting at 3120 mJy beam$^{-1}$, \citealt{Herczeg2017}). $b$: integrated intensity from the molecular emission line of C$^{18}$O 2$-$1 (contours are at 3$\sigma$, 6$\sigma$, 9$\sigma$, etc.). The white crosses mark the position of the SVS~4 sources, and the white star represents SMM~4.}
\label{C18O_depletion_map}
\end{figure*}

\subsection{Combined gas-ice maps}
\label{comb_maps}

In Figure~\ref{ig_maps}, panels \textit{a} and \textit{b} compare the CO ice abundance to the gas-phase $^{13}$CO~2~$-$~1 (panel \textit{a}) and C$^{18}$O~2~$-$~1 (panel \textit{b}) emission. Because of optical depth effects, the C$^{18}$O emission shows more structure than $^{13}$CO and traces the densest regions of the cluster. The interplay between CO ice and CO gas in cold dark regions is mainly dominated by the freeze-out of CO gas onto the dust grains. When we assume that the overall column density of CO molecules (CO gas + CO ice) is constant, an anti-correlation is expected to be observed between CO gas and CO ice. Most of the SVS~4 sources show higher CO ice abundances (e.g. SVS~4$-$6, SVS~4$-$9, and SVS~4$-$10) where the $^{13}$CO emission is more intense, suggesting that part of the variations are due to the overall physical structure of the region. The same relations are seen in panel~\textit{b}. Because ice absorption and gas emission might not be probing the same column of material, this preliminary conclusion has to be verified by comparing ice and gas abundances, and this is addressed in Sect.~\ref{correlation_plots}.

To test whether the C$^{18}$O line emission traces the densest regions of the core, Figure~\ref{C18O_depletion_map} shows a comparison of the 850~$\mu$m dust continuum distribution in SVS~4 with molecular emission traced by C$^{18}$O. It can be observed that the C$^{18}$O emission in SVS 4 does not closely follow the dust emission. However, some care has to be taken in interpreting the dust emission map of SVS~4 because embedded sources can lead to an increase in dust temperature, and consequently, in the submillimeter continuum flux, which alters the dust emission morphology (e.g. \citealt{Chandler1996}). Still, the distributed C$^{18}$O emission does indicate that CO gas is present on larger scales in the outer envelope of SMM~4.

In Figure~\ref{ig_maps}, panels~\textit{c} and \textit{d} compare the CH$_3$OH ice (panel~\textit{c}) and H$_2$O ice (panel~\textit{d}) abundances to the CH$_3$OH~5$_0~-~4_0$~A$^+$ integrated intensity. The peak CH$_3$OH emission likely traces the emission of the outflow associated with SMM~4 \citep{Kristensen2010}: the strongest CH$_3$OH emission is observed in three ridges in the NW/SE direction from the centre of the cluster and not where the majority of the SVS~4 sources are located. Two ridges are localised in the vicinity of SMM~4, and the third is observed in the southern region of the map, in the proximity of the outflow knot position SMM~4$-$S. No straightforward trend is observed between gas-phase methanol and water ice: generally, the H$_2$O ice abundances are higher where the gas-phase methanol emission is stronger (e.g. SVS~4$-$2, SVS~4$-$5, SVS~4$-$6, and SVS~4$-$12). Similarly, the methanol ice abundances are lower where the methanol emission is weaker (e.g. SVS~4$-$4, SVS~4$-$8). However, some deviations from this behaviour are seen for SVS~4$-$2, SVS~4$-$7, and SVS~4$-$10.

It is worth noting that the morphology of the CH$_3$OH emission differs from the emission of the CO isotopologues. While the CH$_3$OH emission is structured, the CO isotopologues show a more uniform distribution. Furthermore, the emission of each molecular tracers is very much dominated by the large-scale structure of SMM~4 and no separate peaks are seen towards any of the SVS~4 sources, suggesting that they do not strongly affect the gas locally.

\begin{figure*}
\centering
\includegraphics[width=5.3in]{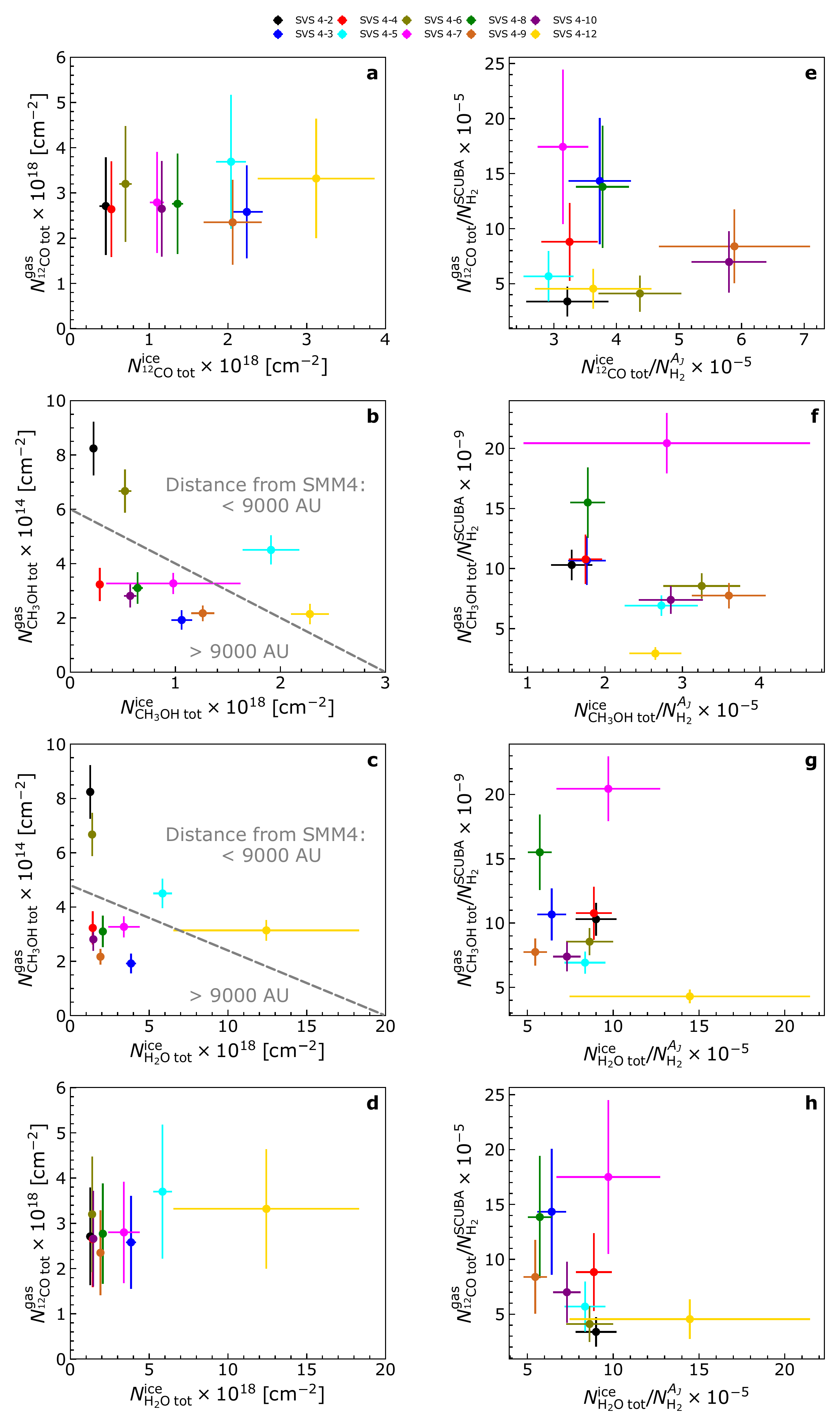}
\caption{Gas and ice variations. Panels {\it a}~$-$~{\it d}: relation between CO, CH$_3$OH, and H$_2$O ice and CO and CH$_3$OH gas column densities. Panels {\it e}~$-$~{\it h}: relation between CO, CH$_3$OH, and H$_2$O ice and CO and CH$_3$OH gas abundances relative to H$_2$. The coloured filled circles identify the SVS~4 stars; the dashed grey line divides the sources into two groups: YSOs located within 9000~AU from SMM~4, and YSOs located at distances >~9000~AU from SMM~4.}
\label{Corr_gas_ice}
\end{figure*}

\begin{figure*}
\centering
\includegraphics[trim={0 0 0 0},clip,width=6.in]{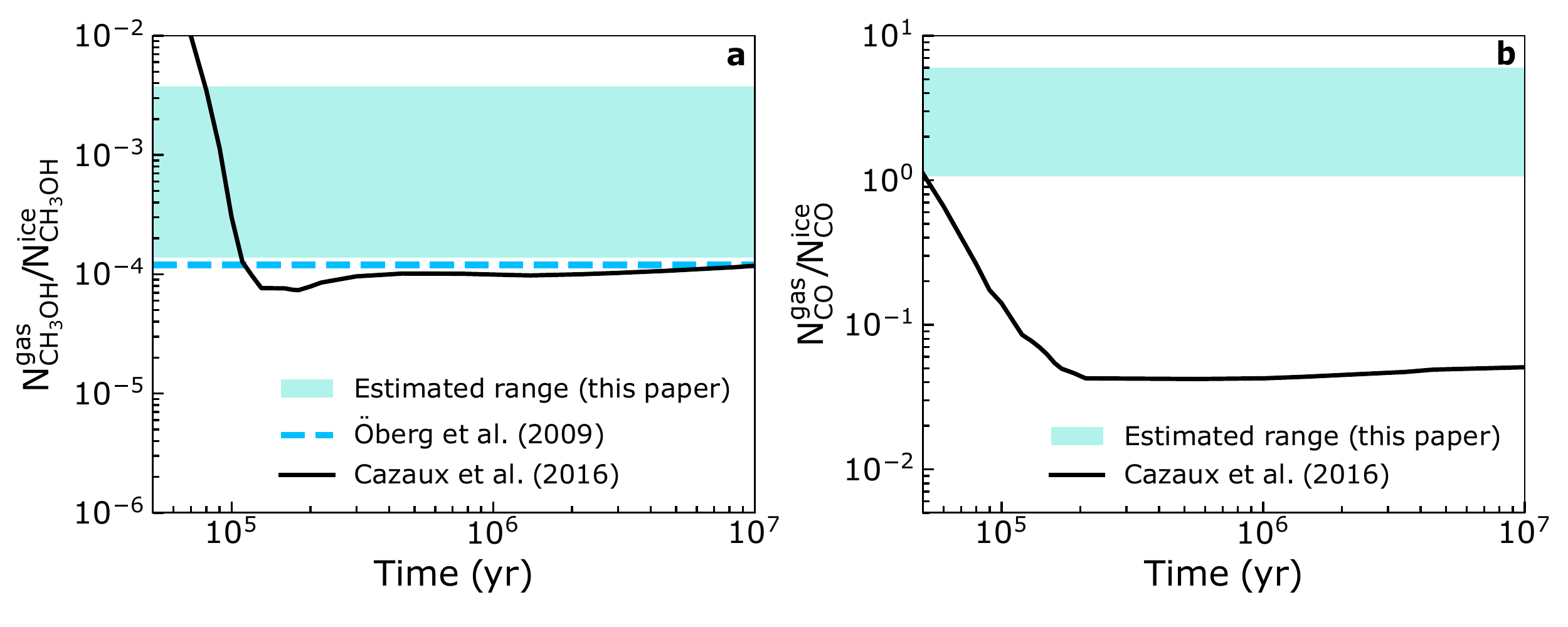}
\caption{CH$_3$OH and CO gas-to-ice ratios ($N_\mathrm{gas}$/$N_\mathrm{ice}$) for the SVS~4 sources in the Serpens molecular core. The shaded azure areas represent the ranges estimated in this paper for both molecules. The curves represented by the solid black lines indicate the theoretical ratio estimated from the astrochemical model by \citet{Cazaux2016}. The dashed light blue line in the left panel marks the ratio estimated by \citet{Oberg2009a}.}
\label{gas_ice_ratio}
\end{figure*}

\subsection{Gas and ice variations}
\label{correlation_plots}

To further explore trends between ice and gas in the region,  Figure~\ref{Corr_gas_ice} compares the inferred values towards each of the SVS~4 cluster members. Figure~\ref{Corr_gas_ice}\textit{a} shows that there is no clear trend between CO ice and $^{12}$CO gas column densities. For instance, the CO gas column density for SVS~4$-$3 and SVS~4$-$4 changes only by 3\%, whereas the ice column density for these sources changes by a factor of 4.3. The $^{12}$CO gas column density is almost uniform at the SVS~4 source positions, suggesting an extended gaseous component that is not sensitive to CO freeze-out.
As discussed in Section~\ref{comb_maps}, an anti-correlation between CO ice and gas is expected in cold dark star-forming regions due to the effect of CO freeze-out. The absence of this trend in SVS~4 indicates that this relation is not valid in this region, which is characterised by a complex physics. The low temperatures in the cluster allow CO freeze-out to occur, but at the same time, due to temperature variations and dust heating around 20~K caused for example by SMM~4 or its outflow, the frozen CO is efficiently released into the gas phase. 

Figure~\ref{Corr_gas_ice}\textit{b} compares the CH$_3$OH gas and ice column densities. The CH$_3$OH gas map in Figure \ref{discussion_plots}{\it b} indicates that the methanol column density is higher around SMM~4 and just south of SVS~4$-$7; it is almost one order of magnitude higher than towards the column density at the SVS~4 source positions. Figure~\ref{Corr_gas_ice}\textit{b} shows that the gas column density spans from $\sim$1.9 to 8~$\times$~10$^{14}$~cm$^{-2}$ and the ice column density from $\sim$0.2 to 2.3~$\times$~10$^{18}$~cm$^{-2}$. The dashed grey line illustrates the sources that are located within 9000~AU from SMM~4. The sources in the vicinity of SMM~4, excluding SVS~4$-$12, show an enhancement in the CH$_3$OH gas column density of a factor of 1.5$-$3.0, in agreement with strong methanol emission associated with the SMM~4 outflow. SVS~4$-$12, a particularly deeply embedded source (A$_\mathrm{J}$=~26~mag), does not follow this trend. On the other hand, no clear enhancement of the methanol ice column densities is seen in the SMM~4 outflow region. It thus appears that the majority of the methanol is in solid form on the grains, even for the LoSs showing gas methanol enhancements.

Fig. \ref{Corr_gas_ice}, panels {\it a} and {\it b, } show that the chemical behaviours of CH$_3$OH and CO are anti-correlated: larger variations are observed for CH$_3$OH gas than for CH$_3$OH ice, while the opposite is seen for CO. This might be partly explained by the higher volatility of CO compared to CH$_3$OH, especially because the temperatures in the cluster are close to the CO sublimation temperature (15$-$20~K).

The CO and CH$_3$OH gas column density are compared to the H$_2$O ice column density in panels {\it c} and {\it d}, although no correlation is expected between them because their formation routes are different. The column density of water ice varies from $\sim$1.3 to 5.9~$\times$~10$^{18}$ cm$^{-2}$ when SVS~4$-$12 is not taken into account. SVS~4$-$12 shows an additional enhancement of a factor of 2. The large error bars for this source are due to the low S/N in the L and M bands and the poor fitting in the red wing between 3.0-3.7~$\mu$m. Such a high variation indicates that this region is cold to keep water in the solid phase, and that simultaneously, the hydrogenation of atomic oxygen to form water occurs efficiently. The same trends observed in panels {\it a} and {\it b} are also seen when CO and methanol gas are compared to H$_2$O ice. 

Figure~\ref{Corr_gas_ice}, panels {\it e} $-$ {\it h,} illustrates the gas and ice abundances towards the cluster. A lack of correlation between the ice and gas species is also observed when the relative abundances are compared. This additional result supports the conclusion that there is no straightforward correlation between CO and CH$_3$OH gas with their ice counterparts in the cluster. This is most likely related to the complexity of the SVS~4 morphology: the presence of the Class~0 protostar and its envelope, the outflow that intersects the cluster, and the vicinity of the SVS~4 stars. All these factors are reflected in temperature and density variations throughout the entire region. 


\section{Discussion}
\label{discussion}

In the previous sections, gas-ice maps have been analysed to investigate the feedback between ice and gas processes. In this section, the information contained in the maps is used to directly determine the CH$_3$OH and CO gas-to-ice ratios ($N_\mathrm{gas}$/$N_\mathrm{ice}$) shown in Figure~\ref{gas_ice_ratio}, calculated from the column densities derived in Sections~\ref{Ice_cd} and \ref{Gas_cd}. The azure shaded areas in the figure indicate the estimated ranges of CH$_3$OH and CO gas-to-ice ratios towards the SVS~4 sources, whereas the dotted light blue line displays the CH$_3$OH gas-to-ice ratio estimated by \citet{Oberg2009a}. The solid black lines show the predictions from the astrochemical model by \citet{Cazaux2016}: the model simulates a typical molecular cloud with a three-phase chemical model that includes gas-phase chemistry that interacts with the surface and bulk chemistry. In the model, thermal desorption and non-thermal desorption of the species in the ice mantles are taken into account. More specifically, two non-thermal desorption mechanisms are included for CO and H$_2$O: (i) absorption of a single UV-photon or induced by H$_2$ ionisation by cosmic rays (i.e. photodesorption), and (ii) desorption promoted by the exothermicity of a chemical reaction (i.e. reactive desorption). The CH$_3$OH ice-to-gas ratio found by \citet{Cazaux2016} agrees very well with the lower value derived in this study. It is important to note that only one non-thermal desorption mechanism for methanol is included in the model, reactive desorption. This keeps the gas-to-ice ratio at about $\mathrm{10^{-4}}$. If additional non-thermal desorption mechanisms were considered, the methanol gas-to-ice ratio calculated by the model would be higher.

The CH$_3$OH gas-to-ice ratio ($N_\mathrm{CH_3OH_{gas}}$/$N_\mathrm{ CH_3OH_{ice}}$) varies between $\mathrm{1.4 \times 10^{-4}}$ and $\mathrm{3.7 \times 10^{-3}}$ (Figure~\ref{gas_ice_ratio}$a$). The higher values correspond to SVS~4$-$2 and SVS~4$-$6, which are located at the CH$_3$OH peak emission close to SMM~4. These two sources also show the highest CH$_3$OH gas column densities. Non-thermal desorption mechanisms might be more efficient in this region of the cluster because of the higher degree of UV irradiation originating from the SMM~4 protostar. 

The lower values reflect the less efficient methanol non-thermal desorption of the protostars located farther away from SMM~4, and that are by inference subjected to a lower degree of UV irradiation. Simultaneously, because the SMM~4 outflow intersects the cluster, gas-phase destruction by shocks (e.g. collisional dissociation) might be responsible for low gas-to-ice ratios in the region. Our derived value is supported by previous measurements by \citet{Oberg2009a} towards four embedded Class 0/I low-mass protostars.

The CO gas-to-ice ratio ($N_\mathrm{CO_{gas}}$/$N_\mathrm{CO_{ice}}$) derived in this study varies between 1 and 6 (Figure~\ref{gas_ice_ratio}$b$). This range is significantly higher than the one predicted in the model by \citet{Cazaux2016}. As discussed previously in Section~\ref{analysis}, according to the DUSTY model calculated for SMM~4 in \citet{Kristensen2010}, the dust temperature in the SVS~4 cluster is estimated to be below 20~K, and at these temperatures, CO is indeed expected to be frozen out on grains. The high relative abundance of gaseous CO might be attributed to the morphology of the SVS~4 cluster, and more specifically, to the extended gaseous component traced in the CO gas observations. Because this component is not included in the model by \citet{Cazaux2016}, this explains the mismatch found in Figure~\ref{gas_ice_ratio}$b$.


\section{Conclusions}
\label{conclusions}
We presented SMA and APEX observations of Serpens SVS~4, a dense cluster of low- to intermediate-mass young stars located in the outer envelope of the Class 0 protostar SMM~4. The SMA observations were combined with APEX single-dish maps and with data from the VLT archive to construct combined gas-ice maps of H$_2$O, $^{12}$CO and CH$_3$OH of the cluster. In these maps the abundances of H$_2$O, $^{12}$CO, and CH$_3$OH ices relative to H$_2$ are displayed with the gas-phase emissions of CH$_3$OH, $^{13}$CO, and C$^{18}$O. 
These suites of maps are powerful probes of the physical and chemical history of dense cores. They allowed us to study the complex relationships between gas and ice, specifically, the processes linking ice to gas-phase density in cold dark regions: freeze-out and non-thermal desorption. Finally, gas-to-ice ratios of CO and CH$_3$OH were directly measured to observationally validate their theoretical predictions. The main conclusions of the paper are summarised below. 

CO gas and H$_2$ do not follow the same density profile. The H$_2$ column density increases from SW towards SMM~4, whereas the CO ridge is located at the centre of SVS~4 cluster. Nevertheless, both distributions are more uniform than that of CH$_3$OH, which shows three strong ridges that are likely associated with the sputtering induced by the SMM~4 outflow. The methanol gas column density in these regions is around one order of magnitude higher that at the SVS~4 source positions.
    
There is no clear trend between CO and CH$_3$OH gas with their ice counterparts in the SVS~4 cluster. This result is expected based on the physical structure of the SVS~4 cluster. Temperature and density variations that affect the chemistry are present in the region, which is strongly influenced by the Class 0 protostar, its outflow, and the high YSO density. The CH$_3$OH and CO chemical behaviours are anti-correlated: larger variations are seen for CH$_3$OH gas than for CH$_3$OH ice, while the opposite is seen for CO.

The CO gas-mapping illustrates the existence of an extended gaseous component that does not probe the densest regions of the cluster where the ice is located. Consequently, the effect of the freeze-out on the CO gas column densities is not observed.This is reflected in higher CO gas column densities and in higher CO gas-to-ice ratios compared to the theoretical ratio estimated by \citet{Cazaux2016}.

The higher values for the CH$_3$OH gas-to-ice ratio, induced by the sources located close to the SMM~4 ridges, agree well with the prediction that methanol ice can be formed on dust grains by surface chemistry and is then released into the gas phase in cold regions through non-thermal desorption. The lower values match the measurements performed by \citet{Oberg2009a} for embedded Class 0/I low-mass protostars. Generally, the methanol gas-to-ice ratio range found in this study promotes a scenario in which the vast majority of the methanol molecules in SVS~4 is in ice form, condensed on the surfaces of dust grains.

Similar studies of other star-forming regions will reveal how universal, for example, the gas-to-ice ratios and trends found for the Serpens SVS~4 cluster are. For example, more isolated young stellar objects are suitable to address how efficiently methanol is desorbed in absence of external irradiation. In contrast, regions dominated by strong external irradiation can constrain how efficient the methanol photodesorption is compared to the other non-thermal desorption mechanisms. 
Future observations with the James Webb Space Telescope (JWST) will overcome the main limitation in the detection of complex molecules in the ice phase, the low spectral resolution. For example, medium-resolution MIRI observations of the 7$-$8~$\mu$m region will allow us to access the ice content of a suite of species more complex than methanol (e.g. CH$_3$CHO and CH$_3$CH$_2$OH). The combination of these data with high-sensitivity ALMA observations will enable us to construct accurate gas-ice maps of several complex organic molecules and to distinguish the interplay between ice and gas in low- and high-mass star-forming regions.

\begin{acknowledgements}
We thank the anonymous referee for the useful comments that improved the manuscript. This work is based on observations with the Submillimeter Array (SMA), Mauna Kea, Hawaii, program code: 2016B-S022; with the Atacama Pathfinder EXperiment (APEX), Llano Chajnantor, Chile, program code: 099.F-9316(A) and with the Very Large Telescope (VLT), Paranal, Chile, under European Southern Observatory (ESO) programmes 075.C-0384(A) and 077.C-0363(A). 
The SubMillimeter Array is a joint project between the Smithsonian Astrophysical Observatory and the Academia Sinica Institute of Astronomy and Astrophysics and is funded by the Smithsonian Institution and the Academia Sinica. The Atacama Pathfinder EXperiment (APEX) telescope is a collaboration between the Max Planck Institute for Radio Astronomy, the European Southern Observatory, and the Onsala Space Observatory. Swedish observations on APEX are supported through Swedish Research Council. The group of JKJ acknowledges the financial support from the European Research Council (ERC) under the European Union's Horizon 2020 research and innovation programme (grant agreement No 646908) through ERC Consolidator Grant "S4F". The research of LEK is supported by a research grant (19127) from VILLUM FONDEN. HJF gratefully acknowledges the support of STFC for Astrochemistry at the OU under grants No ST/P000584/1 and No ST/T005424/1 enabling her participation in this work. GP is grateful to EU COST Action CM1401 "Our Astro-Chemical History" for funding contributions towards the realization of this work. 
\end{acknowledgements}


 \newcommand{\noop}[1]{}



\appendix
\section{Fitting of ice data}
\label{appendixA}

\subsection{Continuum determination}
\label{continuum}

In order to derive the column densities of ice water, carbon monoxide, and methanol, the L- and M-band spectra have to be converted into an optical depth scale given by
\begin{equation}
\tau_{\lambda} = \mathrm{-ln} \left(\frac{F_{\lambda}^\mathrm{obs}}{F_{\lambda}^\mathrm{cont}} \right)
\label{tau}
,\end{equation}
where $F_{\lambda}^{\mathrm{obs}}$ is the observed flux and $F_{\lambda}^{\mathrm{cont}}$ is the continuum spectral energy distribution (SED).

Determining the continuum of embedded sources is not straightforward because several physical parameters of the disc and envelope must be known \citep{Whitney2003a, Gramajo2010, Robitaille2017}, together with accurate photometric measurements. In this paper, the continuum of the L-band observations was derived by the fitting of the near-infrared photometric data in Table~\ref{table:photometry} and the spectral data above 3.6~$\mu$m. The accuracy of the 2MASS photometric data and VLT-ISAAC spectral data is about 8\% and 13\%, respectively. To account for the different flux accuracy, weighted fits were used to determine the continuum: less weight was given to the VLT data compared to 2MASS data. The weights are inversely proportional to the variance of the data. The fitting was performed using one or two blackbodies if $A_J$ towards the source was higher or lower than 9.8 mag, respectively. Although a huge degeneracy involving the disc ($M_{disc}$) and envelope mass ($M_{env}$) and the inclination ($i$) arises in interpreting SED of YSOs, the use of one blackbody to trace the continuum of the L-band observations towards embedded sources ($M_{env} \sim 0.4 M_{\odot}$), as result of the re-emission of the stellar flux by the envelope, is consistent with radiative transfer models \citep{Whitney2003a, Robitaille2006, Robitaille2007} of Class~0 and late Class~0 objects. In less embedded environments, the combination of the scattered or re-emitted photons by the disc and envelope becomes more evident, and a low-infrared excess is observed longward of 3.6~$\mu$m. In this case, a second blackbody component is required to fit the final part of the L band. Two blackbodies were adopted for SVS 4$-$5 because of the high mid-infrared excess, as pointed out by \citet{Pontoppidan2004}, and because SVS 4$-$9 is a peculiar source due to its strong X-ray flux \citep{Preibisch2003}. Because the extinction law at the L bands is significantly lower than the visual extinction (\citealt{Cardelli1989}), we decided to use reddened blackbody functions as also adopted by previous studies \citep{Ishii2002,Chiar2002,Moultaka2004}. The good agreement between the optical depths of the SVS~4 sources obtained in this work and those derived by \citet{Pontoppidan2004} (see Figures 4$-$6 of \citet{Pontoppidan2004} and A.1 and A.2 of this work) contributes to the reliability of this method.
The left-hand sides of Figures \ref{photometry_plot_L} and \ref{photometry_plot_Lp2} show the observed flux, the photometry data, and the synthetic continuum used in this paper. The optical depth, calculated with Equation \ref{tau}, is shown on the right-hand side of both figures. SVS~4$-$12 is a saturated source, and only the red wing is shown. In order to estimate the optical depth, the SVS~4$-$9 curve was scaled to the red wing of SVS~4$-$12, as suggested by \citet{Pontoppidan2004}. The authors stated that the optical depth profile does not change significantly if another SVS 4 source is used in the scaling because the SVS~4 spectra have similar shapes.

\begin{table}
\begin{center}
\small\addtolength{\tabcolsep}{-4pt}
\caption{Photometry of the SVS~4 sources.}
\renewcommand{\arraystretch}{1.1}
\label{table:photometry}
\begin{tabular}{llllll}
\hline \hline
SVS & 1.235~$\mu$m$^{a}$ &1.250~$\mu$m$^{b}$ & 1.635~$\mu$m$^{b}$  & 1.662~$\mu$m$^{a}$    & 2.159~$\mu$m$^{a}$ \\ 
       &    [mJy]   &  [mJy]  &    [mJy]  &    [mJy]       &     [mJy]    \\ \hline      
4$-$2   &  1.59 $\pm$ 0.08  & 1.8 $\pm$ 0.2     & 8.5 $\pm$ 0.9   & 7.73 $\pm$ 0.25  &  16.4 $\pm$ 0.38 \\  
4$-$3   &  $\leq$ 0.09      &  --               & 0.4 $\pm$ 0.1   & $\leq$ 0.24      &  1.59 $\pm$ 0.08 \\   
4$-$4   &  1.37 $\pm$ 0.06  & 1.50 $\pm$ 0.2    & 10.5 $\pm$ 1    & 10.5 $\pm$ 0.3   &  24.5 $\pm$ 0.56 \\ 
4$-$5   &  $\leq$ 0.4       & --                & 0.03$\pm$0.01 & $\leq$ 2.34      &  4.0 $\pm$ 0.3   \\    
4$-$6   &  0.6 $\pm$ 0.06   & 0.5 $\pm$ 0.1     & 3 $\pm$ 0.3     & 3.19 $\pm$ 1.62  &  10.3 $\pm$ 0.49 \\ 
4$-$7   &  $\leq$ 0.08      & 0.015$\pm$0.005 & 0.6 $\pm$ 0.1   & 0.48 $\pm$ 0.08 &  4.3 $\pm$ 0.13  \\  
4$-$8   &  $\leq$ 0.21      & 0.04 $\pm$ 0.01   & 1.8 $\pm$ 0.2   & 1.76 $\pm$ 0.10 &  14.4 $\pm$ 0.37 \\ 
4$-$9   &  $\leq$ 0.38      & 0.21 $\pm$ 0.05   & 9.7 $\pm$ 0.1   & 10.1 $\pm$ 0.39 &  68.6 $\pm$ 2.15 \\  
4$-$10  &  0.79 $\pm$ 0.07 & 0.6 $\pm$ 0.1     & 9.9 $\pm$ 1     & 10.7 $\pm$ 0.44  &  41.4 $\pm$ 1.0  \\  
4$-$12  &  --               & --                & --              & --               &  0.2 $\pm$ 0.02  \\   
\hline 
\end{tabular}

\end{center}
\footnotesize{\textbf{Notes.} $^{a}$ from 2MASS \citep{Skrutskie2006} $^{b}$ from UKRIT \citep{Pontoppidan2004}}.
\end{table}

\begin{figure*}
\centering
\includegraphics[trim={25 100 10 100},clip,width=5in]{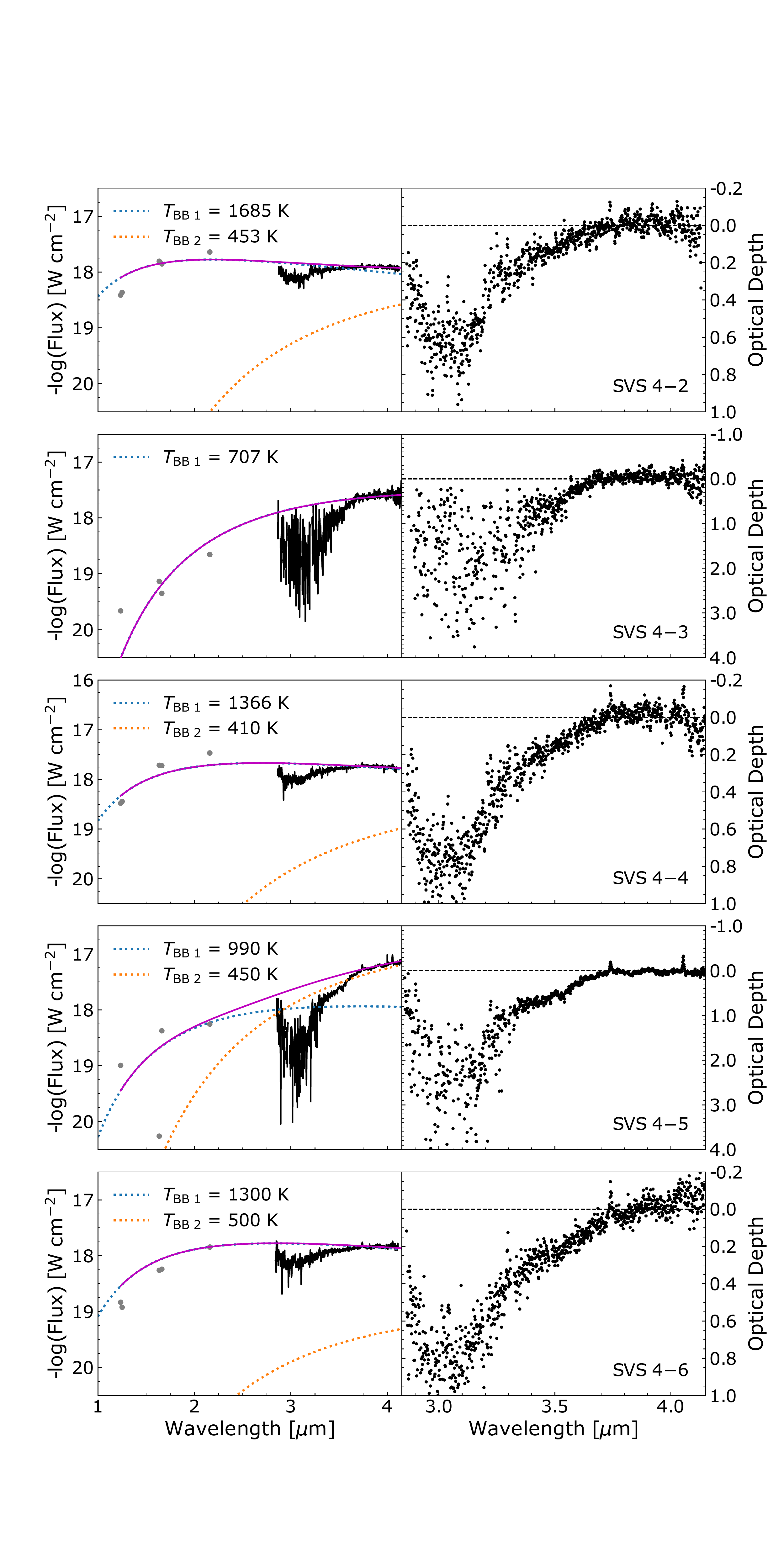}
\caption{L-band SED and optical depth of the sources towards the SVS~4 cluster. \textit{Left:} The dotted blue and orange lines show the blackbody functions we used to determine the continuum (magenta line) by fitting the dark grey photometric points. The temperature of each blackbody component is shown in the upper left corner of each panel and is given in K. \textit{Right:} L-band observations towards the SVS~4 sources on an optical depth scale. The dashed black lines are used as a reference for $\tau$=0.}
\label{photometry_plot_L}
\end{figure*}
  
\begin{figure*}
\centering
\includegraphics[trim={25 100 10 100},clip,width=5in]{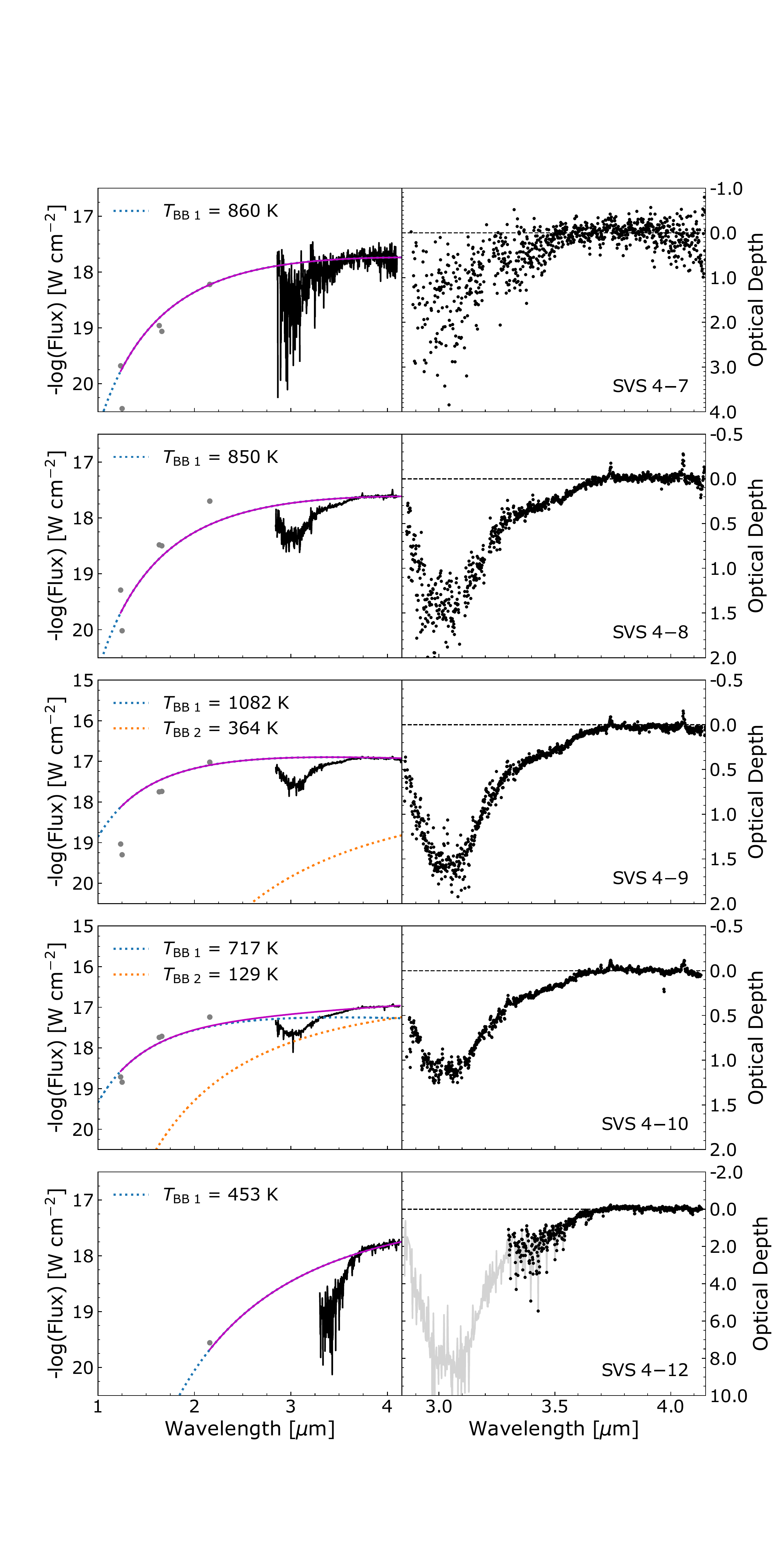}
\caption{Continued from Fig.\ref{photometry_plot_L}. The light grey line represents the SVS 4$-$9 optical depth scaled to the red wing of SVS 4$-$12.}
\label{photometry_plot_Lp2}
\end{figure*}

For the M-band observations, on the other hand, the local continuum was calculated with a spline function. In order to only take the continuum SED into account, the CO and $^{13}$CO~1$-$0 ro-vibrational transitions were masked together with the Pfund HI line. To perform the masking, the CO and $^{13}$CO ro-vibrational spectra obtained from HITRAN database\footnote{https://hitran.org} \citep{Rothman1987} were overlaid on the SEDs. The narrow gas-phase lines were then carefully removed from the SEDs at the designated frequencies. 

\begin{figure*}
\centering
\includegraphics[trim={25 100 10 100},clip,width=5in]{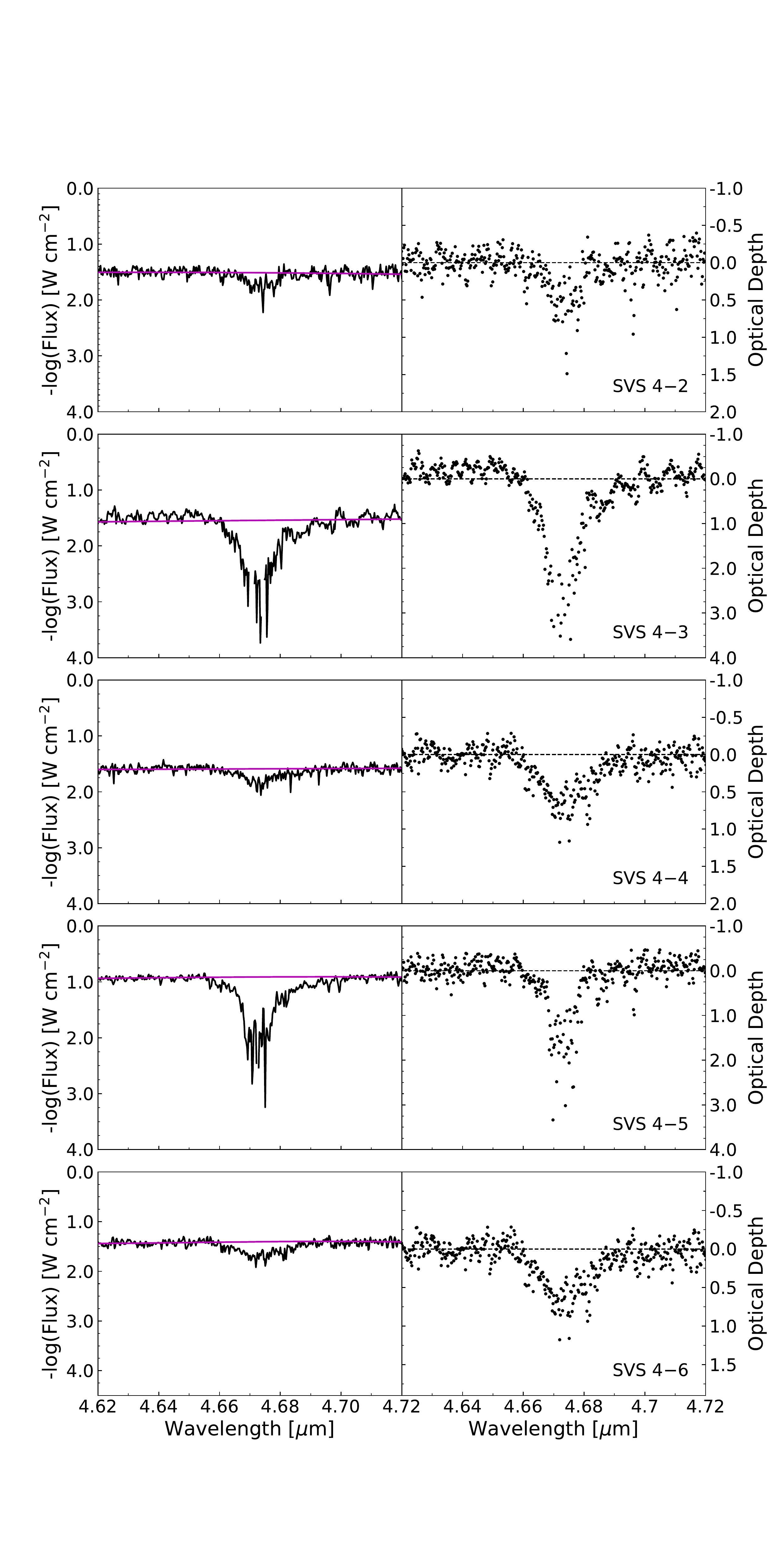}
\caption{M-band SED and optical depth of the sources towards the SVS~4 cluster. \textit{Left:} The spline function we used to determine the continuum for the M-band spectra is shown in magenta. \textit{Right:} M-band observations towards the SVS~4 sources on an optical depth scale. The dashed black lines are used as a reference for $\tau$=0.}
\label{photometry_plot_M}
\end{figure*}
  
\begin{figure*}
\centering
\includegraphics[trim={25 100 10 100},clip,width=5in]{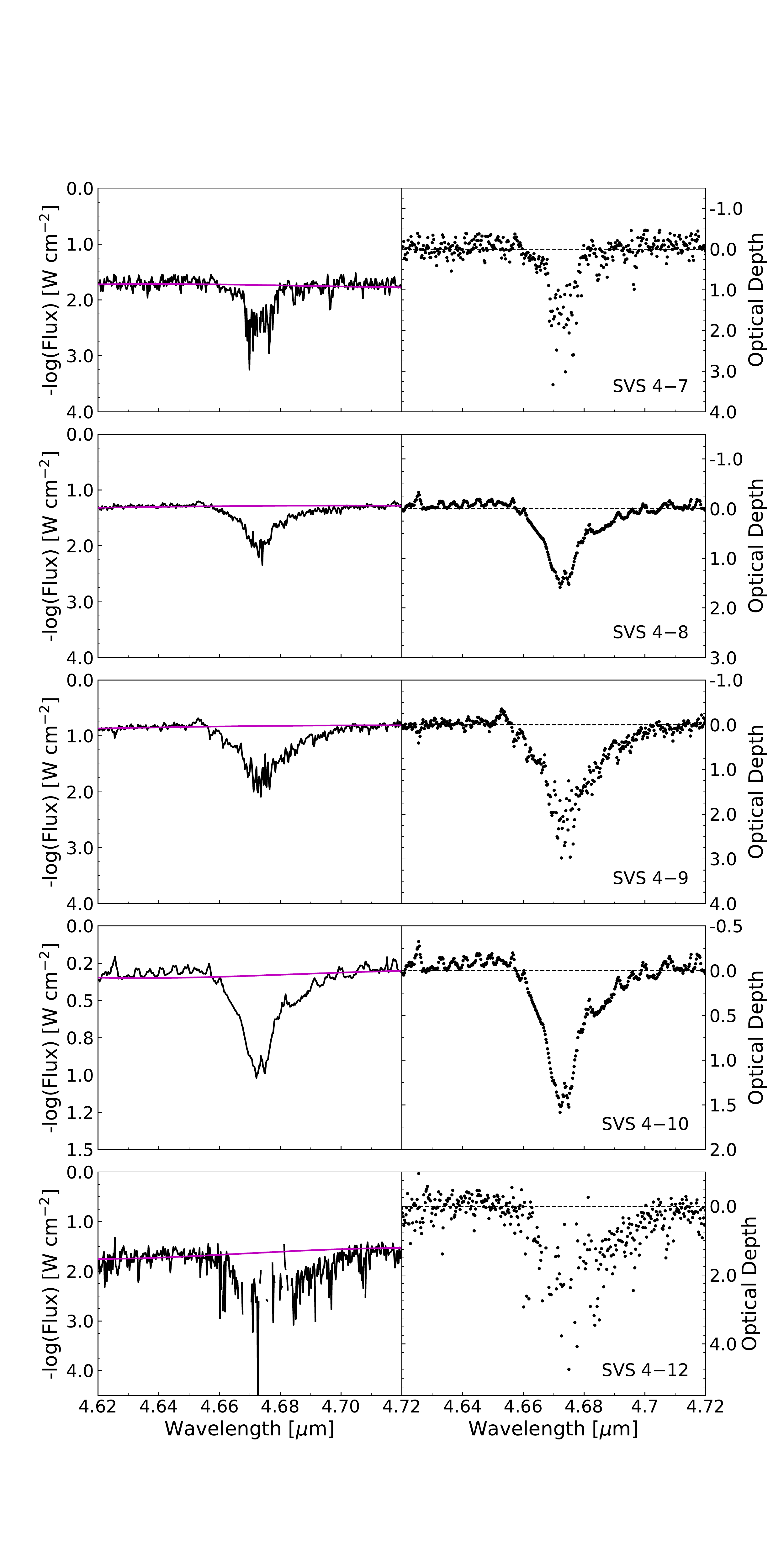}
\caption{Continued from Fig.\ref{photometry_plot_M}.}
\label{photometry_plot_M_p2}
\end{figure*}

\subsection{Optical depth fitting}
\label{Optical depth fitting}

\begin{figure*}
\centering 
\includegraphics[trim={10 0 10 0},clip,width=4.5in]{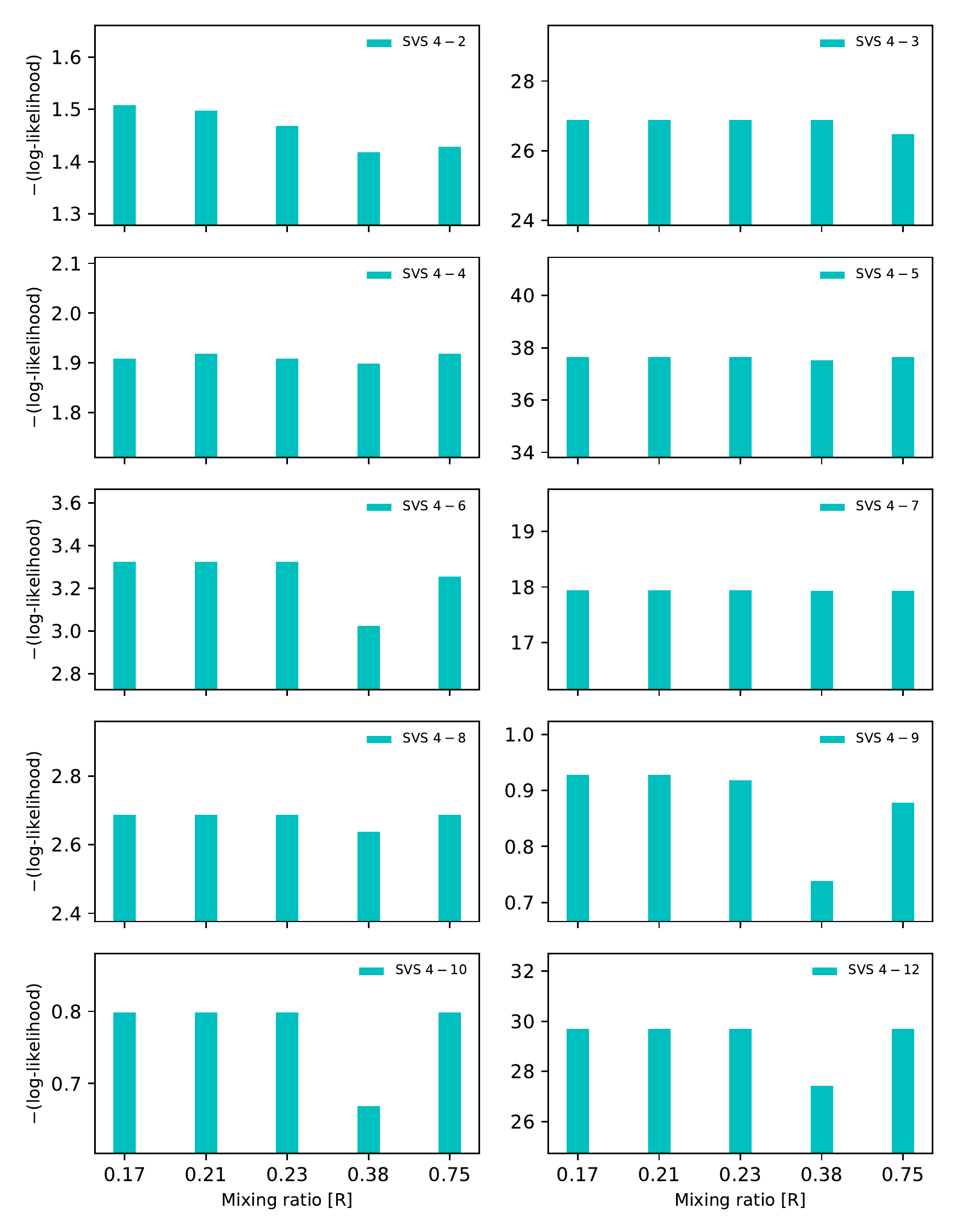}
\caption{Log-likelihood bar plot of the H$_2$O:CH$_3$OH mixing ratios ($R$). The mixture with R=38\% provides a slightly better match of observations and models for all SVS~4 sources, except for SVS~4$-$3.}
\label{MLE}
\end{figure*}

The M-band optical depths were fitted using the OMNIFIT code \citep{aleksi}, an open-source code that aims to fit observational data of astrophysical ices using multi-component laboratory spectra and analytical data. The LMFIT package\footnote{https://lmfit.github.io/lmfit-py/} was used to perform non-linear optimisation with the Levenberg-Marquardt method. M-band optical depths were modelled by combining a CDE-corrected Lorentzian function CO:CH$_3$OH (1:1) laboratory data \citep{Cuppen2011} for the middle and red component, respectively. In the case of the L bands, three laboratory data were used, pure H$_2$O \citep{Fraser2004}, CO:CH$_3$OH (M-band values), and H$_2$O:CH$_3$OH \citep{Dawes2016}. The CO:CH$_3$OH optical depth was kept constant from the M-band fitting. In order to decide the best water-methanol mixing ratio ($R$), five variations were checked and the maximum likelihood estimation (MLE) was used to determine the best model given the higher probability. In order to achieve this goal, the OMNIFIT code was modified and the optimisation procedure adapted to allow the Monte Carlo Markov chain (MCMC) analysis. The log-likelihood function is given by

\begin{equation}
    \mathrm{ln} p(\tau_{\lambda}^{obs}|k \cdot \tau_{\lambda,true}^{lab}) = -\frac{1}{2} \sum_n \left[\frac{(k \cdot \tau_{\lambda,true}^{lab} - \tau_{\lambda}^{obs})^2}{\sigma_n^2} + \mathrm{ln}(2\pi \sigma_n^2)\right]
,\end{equation}
\label{A.2}
where $\tau_{\lambda}^\mathrm{obs}$ was defined in Equation~\ref{tau}, $\tau_{\lambda,true}^\mathrm{lab}$ is the optical depth of the laboratory data, $k$ the scale factor, and $\sigma_n^2$ is the optical depth error, propagated from the flux error. The MLE was initially found using the down-hill simplex method \citep{Nelder1965}, and the best solution used as initial point to sample the posterior distribution using the affine-invariant MCMC package \texttt{emcee} \citep{Foreman2013}. The log-likelihood comparisons are shown in Figure~\ref{MLE}.

\subsection{Derivation of H$_2$O and CH$_3$OH ice column density}
\label{Derivation of H$_2$O and CH$_3$OH ice column density}
The pure H$_2$O column density was calculated by integrating the optical depth of the O-H stretching mode at 3.0 $\mu$m of the fitted laboratory spectra, after decomposing the L-band spectra performed by OMNIFIT, using the equation below:
\begin{equation}
    N_\mathrm{ice} = \frac{1}{\mathcal{A}} \int_{\nu_1}^{\nu_2} \tau_{\nu} d\nu
    \label{col_dens}
,\end{equation}
where $\mathcal{A}$ is the band strength of a specific vibrational mode, and $\nu$ must be understood as the wavenumber in units of cm$^{-1}$. The band strength we used was 2 $\times$ 10$^{-16}$ cm molecule$^{-1}$ taken from \citet{Hudgins1993}.

Because we lack accurate band strengths for the H$_2$O:CH$_3$OH component, mixed water and methanol column densities were derived from non-saturated and negligibly blended bands at 1660 cm$^{-1}$ (6.0 $\mu$m) and 2827 cm$^{-1}$ (3.57 $\mu$m), respectively, following the empirical relations found by \citet{Giuliano2016} for a binary mixture of H$_2$O:CH$_3$OH:

\begin{equation}
    N_\mathrm{H_2O} = (29067 \times \tau_{1660 \; \mathrm{cm^{-1}}} - 300) \times 10^{15} \; \mathrm{molec \; cm^{-2}}
\end{equation}
\begin{equation}
    N_\mathrm{CH_3OH} = (3804 \times \tau_{2827 \; \mathrm{cm^{-1}}} + 4) \times 10^{15} \; \mathrm{molec \; cm^{-2}}
,\end{equation}
where $\tau_{1660 \; \mathrm{cm^{-1}}}$ and $\tau_{2827 \; \mathrm{cm^{-1}}}$ are the optical depth values at this position in the spectrum.

\subsection{Derivation of CO ice column density}
\label{Derivation of CO ice column density}
The pure CO ice column density was calculated by the equation below, taken from \citet{Pontoppidan2003b}:
\begin{equation}
    N_\mathrm{Pure \; CO} = 6.03 \; \mathrm{cm^{-1}} \times \tau_{\mathrm{peak}} \times A_{\mathrm{bulk}}^{-1} 
,\end{equation}
where $A_{\mathrm{bulk}}^{-1}$ is the band strength of the bulk (1.1 $\times$ 10$^{-17}$ cm molecule$^{-1}$ $-$ \citealt{Gerakines1995}), and the numerical factor takes the effect of the CDE correction into account.

The column density of CO trapped in the CO:CH$_3$OH sample was calculated from the oscillator number method, described in \citet{Suutarinen2015}. A modified version of Equation \ref{col_dens} was used,
\begin{equation}
\label{A.7}
    N = k \cdot \frac{1}{\mathcal{A}} \int_{\nu_1}^{\nu_2} \tau_{\nu} d\nu
,\end{equation}
where $k$ is a scale factor calculated by OMNIFIT (See Equation~\ref{A.2}), and the oscillator number is the entire term given by $\frac{1}{\mathcal{A}} \int_{\nu_1}^{\nu_2} \tau_{\nu} d\nu$. The value used for the oscillator number of CO and CH$_3$OH was 5.183 $\times$ 10$^{17}$ cm$^{-2}$ taken from \citet{Suutarinen2015}.

\section{Producing gas-phase maps} 
\label{appendixB}
\subsection{Combination of interferometric and single-dish data} 
\label{combination}
The combination of interferometric and single-dish data was performed using CASA version 4.6.0 and following the CASA Guide\footnote{https://casaguides.nrao.edu/index.php} M100 Band3 Combine 4.3 (for a detailed explanation of the combination, see the CASA Guide and references therein).
Initially, the APEX dataset was resampled to match the shape of the SMA datacube using the {\fontfamily{qcr}\selectfont imregrid} task in CASA. Subsequently, the two datasets were trimmed with the CASA {\fontfamily{qcr}\selectfont imsubimage} task to exclude the noisy edge regions in the single-dish data and the regions masked by the {\fontfamily{qcr}\selectfont clean} task in the interferometric data. Then the SMA non-uniform primary beam response was multiplied to the single-dish data by applying the CASA task {\fontfamily{qcr}\selectfont immath}. This step was performed to ensure that the two datasets had a common response on the sky. The SMA and APEX datasets were then combined by executing the CASA task  {\fontfamily{qcr}\selectfont feather}. The {\fontfamily{qcr}\selectfont feather} algorithm applies a Fourier transform and scales the single-dish (low-resolution) data to the interferometric (high-resolution) data. Finally, it merges the two datasets with different spatial resolution, and in our case, it folds into the SMA data the short-spacing corrections of the APEX telescope. 

Figures \ref{mom0_meth}-\ref{mom0_c18o} show the integrated intensity maps of the CH$_3$OH 5$_0~-~4_0$~A$^+$ line and of the CO isotopologues ($^{13}$CO and C$^{18}$O) lines using the SMA data, APEX data, and the combined SMA~+~APEX data. The maps distinctly illustrate that the SMA data filter out spatially extended emission associated with the cloud. On the other hand, the extended emission is very well recovered in the single-dish APEX data. Therefore it is crucial to combine interferometric and large-scale single-dish data when a quantitative analysis of the datasets is performed (e.g. derivation of column densities). 
The species for which most of the flux is recovered in the combination is $^{13}$CO, followed by C$^{18}$O and CH$_3$OH. 

\subsection{CO isotopologue optical depth}
\label{line_ratio}
\begin{figure}
\centering
\includegraphics[width=2.6in]{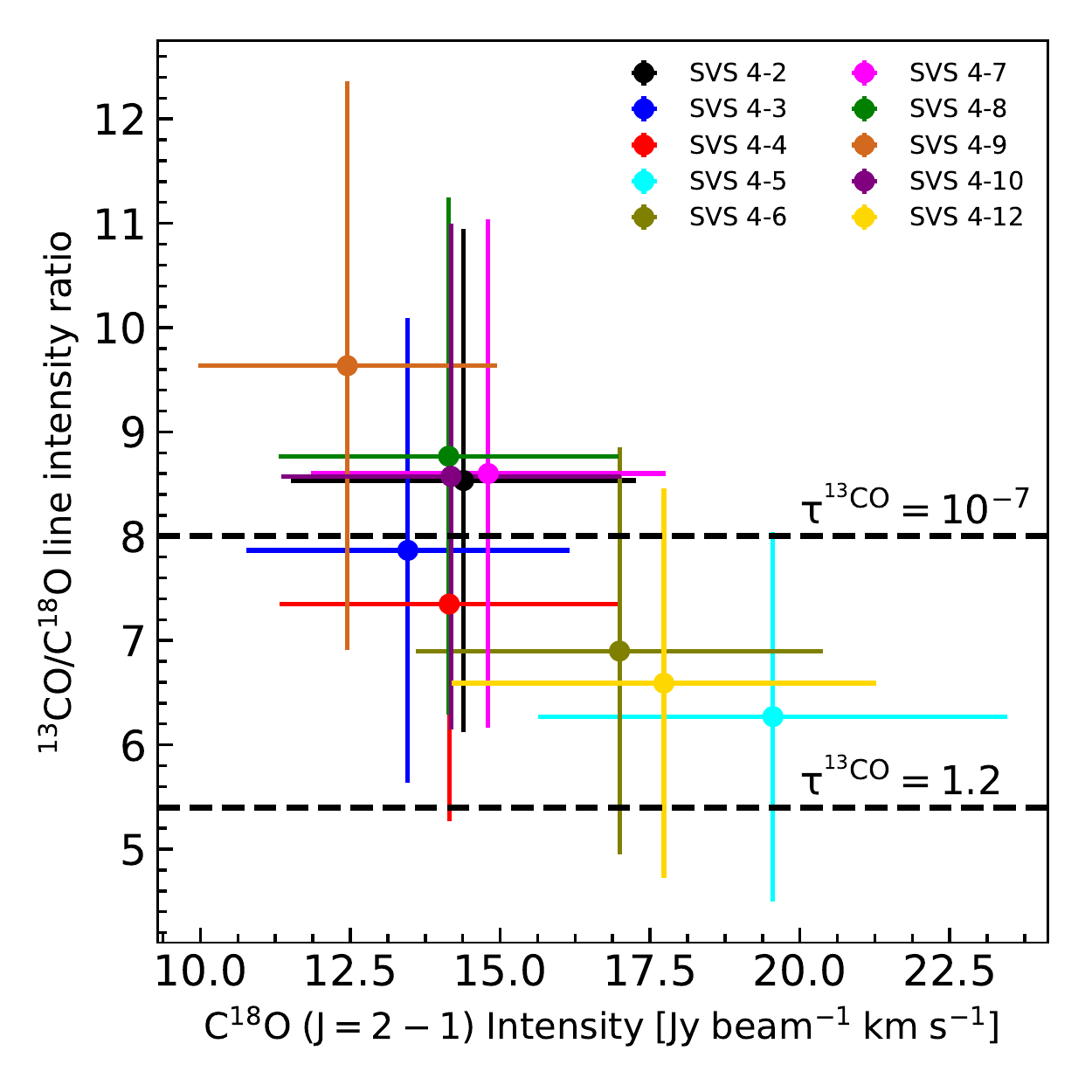}
\caption{$I(\mathrm{^{13}CO})$/$I(\mathrm{C^{18}O})$ line intensity ratio vs.{\it } the $I(\mathrm{C^{18}O})$ line intensity. The dashed lines indicate two optical depth regimes for $\mathrm{^{13}CO}$; the sources are represented by coloured dots. The error bars assume a 20\% error from the flux calibration.}
\label{c18ovs13co}
\end{figure}

\begin{figure*}
  \centering
  \includegraphics[width=7.in]{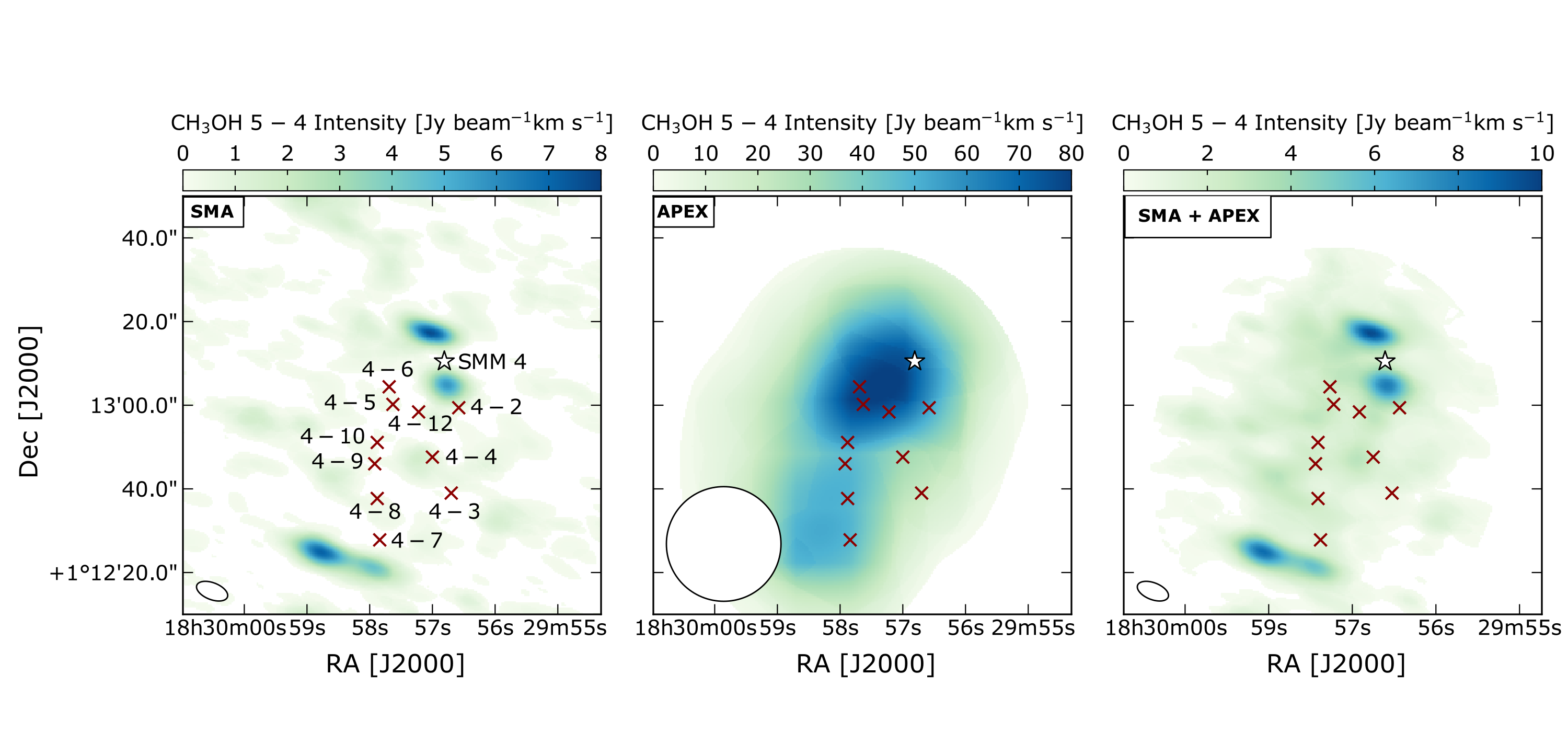}
      \caption{\textit{Left:} CH$_3$OH 5$_0$ $-$ 4~$_0$~A$^+$ line emission observed by the SMA. \textit{Middle:} CH$_3$OH 5$_0$ $-$ 4$_0$~A$^+$ line emission detected by the APEX telescope. \textit{Right:} CH$_3$OH 5$_0$ $-$ 4$_0$~A$^+$ line emission resulting from the SMA~+~APEX combination. All lines are integrated between 5 and 13~km~s$^{-1}$. Contours are at 3$\sigma$, 6$\sigma$, 9$\sigma$, etc. The synthesised beam is shown in white in the bottom left corner of each panel. The dark red crosses and the white star illustrate the positions of the SVS~4 stars and of SMM~4, respectively.}
         \label{mom0_meth}
\end{figure*}

\begin{figure*}
  \centering
  \includegraphics[width=7.in]{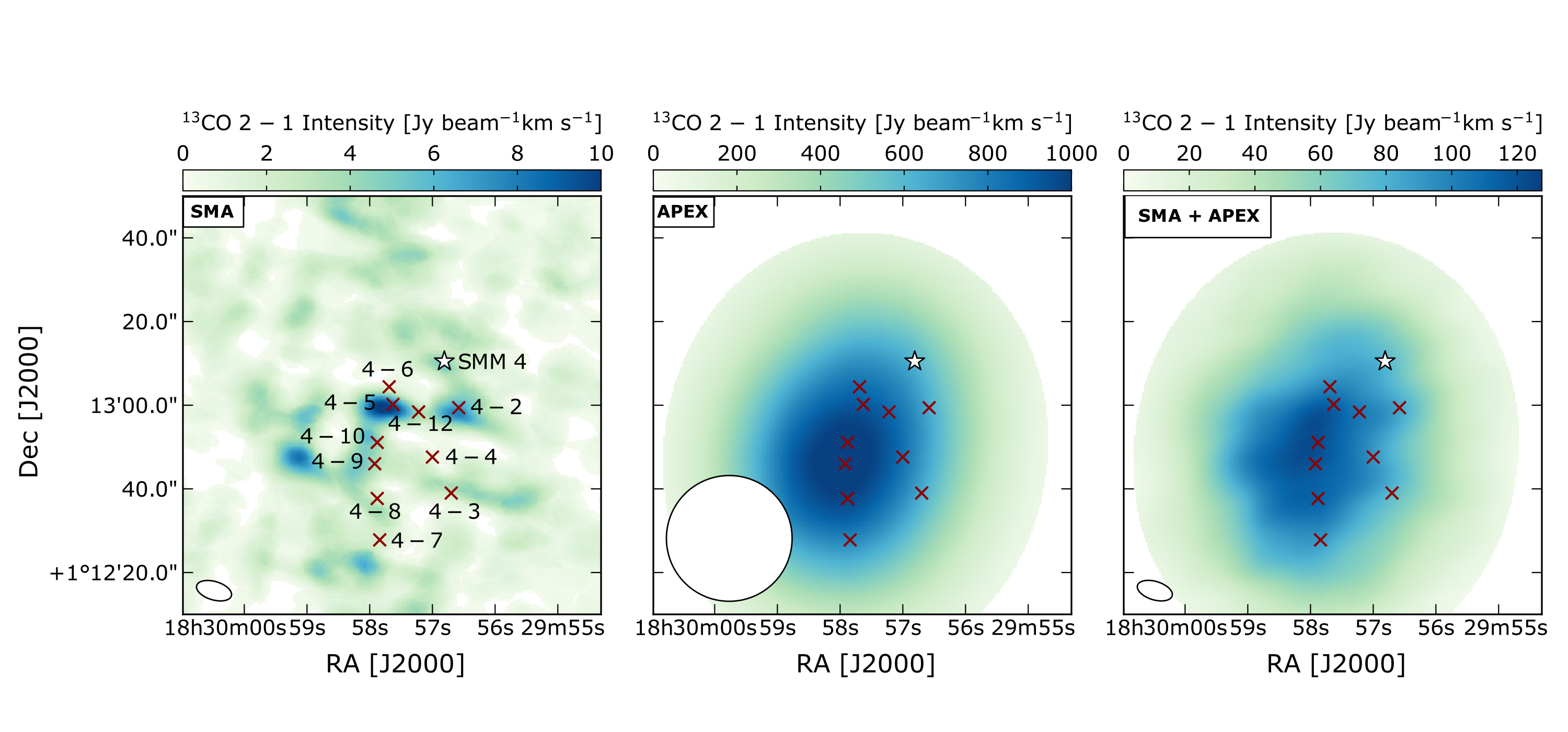}
      \caption{\textit{Left:} $^{13}$CO 2$-$1 line emission observed by the SMA. \textit{Middle:} $^{13}$CO 2$-$1 line emission detected by the APEX telescope integrated between 0 and 15~km~s$^{-1}$. \textit{Right:} $^{13}$CO 2$-$1 line emission resulting from the SMA~+~APEX combination. All lines are integrated between 0 and 13~km~s$^{-1}$. Contours are at 3$\sigma$, 6$\sigma$, 9$\sigma$, etc. The synthesised beam is shown in white in the bottom left corner of each panel. The dark red crosses and the white star illustrate the positions of the SVS~4 stars and of SMM~4, respectively.}
         \label{mom0_13co}
\end{figure*}

\begin{figure*}
  \centering
  \includegraphics[width=7.in]{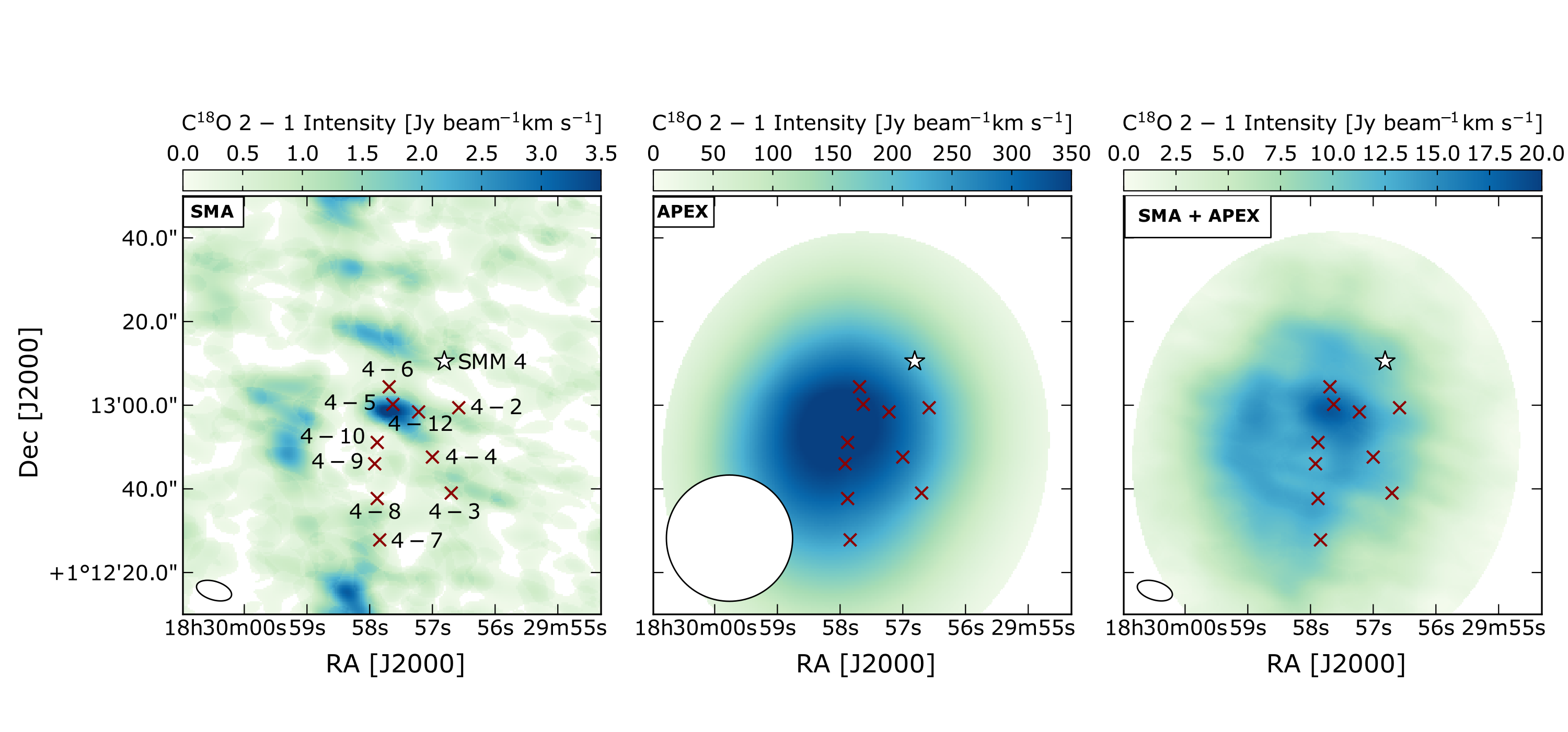}
      \caption{\textit{Left:} C$^{18}$O 2$-$1 line emission observed by the SMA. \textit{Middle:} C$^{18}$O 2$-$1 line emission detected by the APEX telescope. \textit{Right:} C$^{18}$O 2$-$1 line emission resulting from the SMA~+~APEX combination. All the lines are integrated between 4 and 11~km~s$^{-1}$. Contours are at 3$\sigma$, 6$\sigma$, 9$\sigma$, etc. The synthesised beam is shown in white in the bottom left corner of each panel. The dark red crosses and the white star illustrate the position of the SVS~4 stars and of SMM~4, respectively.}
         \label{mom0_c18o}
\end{figure*}

When the gas ratio abundance $\mathrm{^{12}CO}$:$\mathrm{^{13}CO}$:$\mathrm{C^{18}O}$ is constant, the $^{13}$CO optical depth can be calculated from the line intensity ratio \citep{Myers1983, Ladd1998}, given by

\begin{equation}
    \frac{I(\mathrm{^{13}CO})}{I(\mathrm{C^{18}O})} \approx \frac{1 - \mathrm{exp}(-\tau^{\mathrm{^{13}CO}})}{1 - \mathrm{exp}(-\tau^{\mathrm{^{13}CO}}/f)}
,\end{equation}
where $I(\mathrm{^{13}CO})$ and $I(\mathrm{C^{18}O})$ are the line intensities of $\mathrm{^{13}CO}$ and $\mathrm{C^{18}O}$, respectively. $\tau^{\mathrm{^{13}CO}}$ is the $\mathrm{^{13}CO}$ optical depth, and $f$ is the $\mathrm{^{13}CO}/\mathrm{C^{18}O}$ abundance ratio. In the literature, the intrinsic ratio for the Milky Way disc is about 7$-$10 \citep{Wilson1992, Barnes2015}. In this paper, $f$ = 8 is assumed. Figure~\ref{c18ovs13co} shows the $I(\mathrm{^{13}CO})$/$I(\mathrm{C^{18}O})$ line intensity ratio {\it \textup{versus}} the $I(\mathrm{C^{18}O})$ line intensity, where the sources are identified by the colours. The two dashed lines indicate the line ratio associated with the $\tau^{\mathrm{^{13}CO}}$. In the trivial case that the $\mathrm{^{13}CO}$ emission is optically thick and the $\mathrm{C^{18}O}$ emission is optically thin, $\mathrm{C^{18}O}$ increases compared to $\mathrm{^{13}CO}$ along the LoS. Consequently, the ratio between the two lines will decrease, as shown in Figure~\ref{c18ovs13co}. Although the $\tau^{\mathrm{^{13}CO}}$ does indicate the $\mathrm{^{13}CO}$ emission is only moderately optically thick, the decreasing profile indicates that the $\mathrm{C^{18}O}$ emission is optically thin and is therefore more reliable to calculate the $\mathrm{^{12}CO}$ column density assuming local thermodynamic equilibrium (LTE) conditions. This result is still valid when an $\mathrm{^{13}CO}/\mathrm{C^{18}O}$ abundance ratio ($f$) equal to 7 or 10 is assumed. 

\begin{table}
\begin{center}
\caption{Integrated $^{13}$CO and C$^{18}$O line intensities in units of Jy~beam$^{-1}$~km~s$^{-1}$ over each source position.}
\label{table:co_iso_intensities}
\renewcommand{\arraystretch}{1.}
\begin{tabular}{l c c c c }
\hline \hline
Source & $^{13}$CO ($J = 2-1$)  & C$^{18}$O ($J = 2-1$)\\
\hline
SVS~4$-$2   & 122.80 $\pm$ 24.55 &  14.39 $\pm$ 2.86 & \\
SVS~4$-$3   & 105.36 $\pm$ 21.13 &  13.46 $\pm$ 2.73 & \\
SVS~4$-$4   & 104.00 $\pm$ 20.72 &  14.15 $\pm$ 2.80 & \\
SVS~4$-$5   & 122.59 $\pm$ 24.52 &  19.55 $\pm$ 3.91 & \\
SVS~4$-$6   & 117.22 $\pm$ 23.48 &  16.99 $\pm$ 3.38 & \\
SVS~4$-$7   & 127.30 $\pm$ 25.46 &  14.80 $\pm$ 2.96 & \\
SVS~4$-$8   & 123.97 $\pm$ 24.98 &  14.14 $\pm$ 2.93 & \\
SVS~4$-$9   & 119.95 $\pm$ 24.00 &  12.45 $\pm$ 2.50 & \\
SVS~4$-$10  & 121.57 $\pm$ 24.32 &  14.18 $\pm$ 2.81 & \\
SVS~4$-$12  & 116.87 $\pm$ 23.41 &  17.73 $\pm$ 3.51 & \\ 
\hline \hline
\end{tabular}
\end{center}
\begin{center}
\end{center}
\end{table}

\begin{table*}
\begin{center}
\caption{Integrated CH$_3$OH line intensities in units of Jy~beam$^{-1}$~km~s$^{-1}$ over each source position.}
\label{table:spectral_data2}
\renewcommand{\arraystretch}{1.1}
\begin{tabular}{l c c c c c}
\hline \hline
SVS & 5$_0~-~4_0$  E$^+$& 5$_1~-~4_1$  E$^-$& 5$_0~-~4_0$  A$^+$& 5$_1~-~4_1$  E$^+$& 5$_2~-~4_2$  E$^-$ \\
\hline
SVS~4$-$2   & 0.66 $\pm$ 0.13 &  6.23 $\pm$ 1.24 &  7.98 $\pm$ 1.60 &  0.50 $\pm$ 0.10 &  0.19 $\pm$ 0.10  \\
SVS~4$-$3   & ... & 0.69 $\pm$ 0.14 & 0.97 $\pm$ 0.20 & ... & ...                                          \\
SVS~4$-$4   & ... & 1.73 $\pm$ 0.35 & 1.10 $\pm$ 0.22 & ... & ...                                          \\
SVS~4$-$5   & 0.63 $\pm$ 0.13 &  3.82 $\pm$ 0.76 &  1.65 $\pm$ 0.50 &  0.41 $\pm$ 0.13 &  0.12 $\pm$ 0.05  \\
SVS~4$-$6   & 0.50 $\pm$ 0.10 &  3.37 $\pm$ 0.67 &  3.89 $\pm$ 0.78 &  0.85 $\pm$ 0.17 &  0.92 $\pm$ 0.19  \\
SVS~4$-$7   & 0.16 $\pm$ 0.05 &  2.74 $\pm$ 0.60 &  1.26 $\pm$ 0.45 &  0.43 $\pm$ 0.10 &  0.33 $\pm$ 0.07  \\
SVS~4$-$8   & ... & 1.32 $\pm$ 0.26 &  1.32 $\pm$ 0.26 & ... & ...                                         \\
SVS~4$-$9   & 0.30 $\pm$ 0.07 &  1.48 $\pm$ 0.30 &  1.97 $\pm$ 0.40 &  0.12 $\pm$ 0.04 & ...               \\
SVS~4$-$10  & 0.32 $\pm$ 0.07 &  1.12 $\pm$ 0.22 &  2.42 $\pm$ 0.48 & ... & ...                            \\
SVS~4$-$12  & 0.11 $\pm$ 0.04 &  4.29 $\pm$ 0.86 &  4.55 $\pm$ 0.91 &  0.19 $\pm$ 0.05 &  0.25 $\pm$ 0.07  \\ 
\hline \hline
\end{tabular}
\end{center}

\begin{center}
\begin{tablenotes}
\small
\item{\textbf{Notes.} Not provided numbers mean that the integrated intensities are below 5$\sigma$ of the noise level.}
\end{tablenotes}
\end{center}
\end{table*}

\section{Producing an H$_2$ column density map}
\label{appendixC}
The production of the H$_2$ column density map is accomplished by making use of the submillimeter continuum (SCUBA-2) maps at 850 $\mu$m \citep{Herczeg2017}. The equations and the conversion factors used here come from \citet{Kauffmann2008} and are briefly explained below.  

The H$_2$ column density $N_\mathrm{H_2}$ is related to the observed flux per beam $S^\mathrm{beam}_{\nu}$, the dust opacity $k_{\nu}$ , and the Planck function $B_{\nu}$($T$) by
\begin{equation}
\label{C1}
N_\mathrm{H_2} = \frac{S^\mathrm{beam}_{\nu}}{\Omega_\mathrm{A}~\mu_\mathrm{H_2}~m_\mathrm{H}~k_{\nu}~B_{\nu}(T)} 
,\end{equation}

\noindent where $\Omega_\mathrm{A}$ is the beam solid angle, $\mu_\mathrm{H_2}$ is the molecular weight per hydrogen molecule, and $m_\mathrm{H}$ is the hydrogen atom mass. When rearranged in useful units, Equation~\ref{C1} yields 

\begin{equation}
\begin{split}
N_\mathrm{H_2} = 2.02~\times~10^{20}~\mathrm{cm}^{-2} (e^{1.439~(\lambda \mathrm{/ mm)}^{-1} (T/10~\mathrm K)^{-1}}-1) \\ \left( \frac{k_{\nu}}{0.01~\mathrm{cm^2~g}^{-1}} \right) \left( \frac{S^\mathrm{beam}_{\nu}}{\mathrm{mJy~beam}^{-1}} \right)^{-1} \left( \frac{\theta_\mathrm{HPBW}}{10~\mathrm{arcsec}} \right)^{-2} \left( \frac{\lambda}{\mathrm {mm}}\right)^{-3} 
\end{split}
,\end{equation}

\noindent where the dust opacity $k_{\nu}$ is 0.0182~cm$^2$~g$^{-1}$, the telescope beam $\theta_\mathrm{HPBW}$ is 15~arcsec, and the wavelength of the observations $\lambda$ is 850~$\mu$m.

The dust opacity adopted in the calculation of the H$_2$ column density comes from \citet{Ossenkopf1994}, where an H-density of 10$^6$~cm$^{-3}$ and dust with thin ice-mantles coagulating for 10$^5$yr were assumed. The adopted dust temperatures are 10, 15, and 20~K. The dust opacities and dust temperature comply with the standard assumptions made by the c2d collaboration. The dust emission is provided by ground-based sub-millimeter observations at 850~$\mu$m from the SCUBA-2 camera at the James Clerck Maxwell Telescope as part of the proposal ID M16AL001 \citep{Herczeg2017}.

\begin{figure}
\centering
\includegraphics[trim={10 10 10 30},clip,width=2.8in]{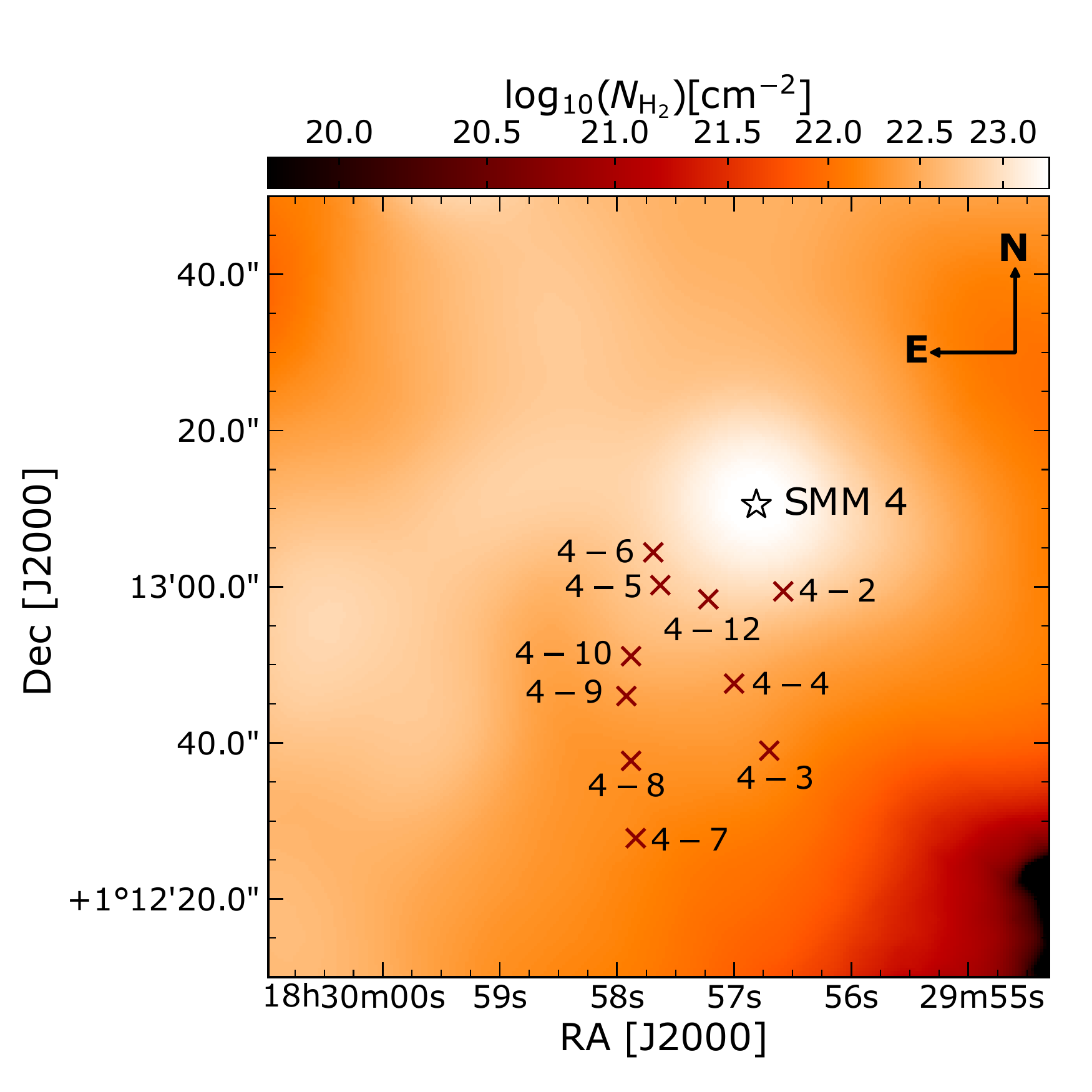}
\caption{H$_2$ column density map of  SVS~4, calculated from the SCUBA-2 dust emission map at 850 $\mu$m from \citet{Herczeg2017}, assuming $T_\mathrm{dust}$~=~15~K. The crosses indicate the position of the SVS~4 stars and of SMM~4.}
\label{NH2_map}
\end{figure}

\end{document}